\documentclass[aps,pra,amsmath,amssymb,reprint,floatfix,superscriptaddress,showpacs]{revtex4-1}
\usepackage{graphicx}
\usepackage[bookmarks=false,colorlinks=true,linkcolor=blue,filecolor=blue,citecolor=blue,urlcolor=blue]{hyperref}
\usepackage{bm}
\usepackage{soul}

\newcommand{\id}{{\bm 1}}

\newcommand{\bfa}{{\bm a}}

\newcommand{\tid}{\tilde{\id}}
\newcommand{\tq}{\tilde{q}}
\newcommand{\te}{\tilde{e}}
\newcommand{\tm}{\tilde{m}}
\newcommand{\tpsi}{\tilde{\psi}}
\newcommand{\tplus}{\widetilde{+}}
\newcommand{\tminus}{\widetilde{-}}
\newcommand{\tpm}{\widetilde{\pm}}
\newcommand{\tmp}{\widetilde{\mp}}
\newcommand{\tA}{\widetilde{A}}
\newcommand{\tB}{\widetilde{B}}
\newcommand{\tL}{\widetilde{L}}
\newcommand{\tM}{\widetilde{M}}
\newcommand{\tX}{\widetilde{X}}
\newcommand{\tZ}{\widetilde{Z}}
\newcommand{\tTC}{\widetilde{\textrm{TC}}}

\newcommand{\sfv}{\mathsf{v}}
\newcommand{\sfp}{\mathsf{p}}
\newcommand{\sfx}{\mathsf{x}}
\newcommand{\sfy}{\mathsf{y}}
\newcommand{\sfz}{\mathsf{z}}
\newcommand{\sfA}{\mathsf{A}}
\newcommand{\sfB}{\mathsf{B}}

\begin{document}

\title{Anisotropic layer construction of anisotropic fracton models}

\author{Yohei Fuji}
\affiliation{Department of Applied Physics, University of Tokyo, Tokyo 113-8656, Japan}

\date{\today}

\begin{abstract}
We propose a coupled-layer construction of a class of fracton topological orders in three spatial dimensions, which  has no immobile excitations but is characterized by single quasiparticle excitations constrained in one-dimensional subspaces and dipole excitations mobile in two-dimensional subspaces. 
The simplest model is obtained by stacking and coupling layers of the two-dimensional toric codes on the square lattice and can be exactly solved in the strong-coupling limit. 
The resulting subdimensional excitations are understood as a consequence of anyon pair condensation induced by the coupling between layers. 
We also present generalizations of the construction for layers of the Kitaev-honeycomb models, the $Z_N$ toric codes, and the toric codes and the doubled semion models on the honeycomb lattice.
\end{abstract}

\maketitle

\tableofcontents

\section{Introduction}

A peculiar phenomenon in strongly interacting many-body quantum systems is the emergence of fractionalized quasiparticles as low-energy collective excitations.
In two dimensions, topologically ordered states, such as the fractional quantum Hall states, support point-like quasiparticles with nontrivial braiding statistics of neither fermion nor boson \cite{Wen16}. 
In three dimensions, fractionalized quasiparticles appear in the shape of point or loop and can have nontrivial statistics among them \cite{Hamma05, ChenjieWang14, SJiang14, CHLin15, JCWang15}. 
These topologically ordered phases are characterized by quasiparticles deconfined in the full two-dimensional ($2d$) or three-dimensional ($3d$) space, and their presence results in a finite ground-state degeneracy on a torus or other nontrivial closed manifolds, which is robust against any local perturbations.

Recently, a new class of topological phases of matter in three dimensions, dubbed fracton topological order \cite{Vijay16}, has been discovered and offers a rapidly growing field of theoretical research \cite{Chamon05, Bravyi11, Haah11, Yoshida13, Vijay15, Vijay16, Williamson16, Halasz17, HMa17, Petrova17, Prem17, Pretko17a, Pretko17b, Pretko17d, Slagle17a, Slagle17b, THHsieh17, Vijay17a, Vijay17b, BShi18, Bulmash18a, Bulmash18b, Devakul18b, Gromov18, HMa18a, HMa18b, HHe18, Pai18, Prem18a, Prem18b, Pretko18b, Pretko18c, Schmitz18b, Shirley18a, Shirley18b, Slagle18b, Williamson18, YYou18a, YYou18b, YYou18c, Bulmash19a, Bulmash19b, Dua19, Gromov19, HYan19a, HYan19b, HSong19, Tian19, Pai19, Prem19a, Prem19b, Shirley19a, Shirley19c, Shirley19d, Slagle19, Sous19, TWang19, YYou19b}; see also a review \cite{Nandkishore19}. 
Quasiparticle excitations emerging from such fracton topological phases are completely immobile or mobile only within lower-dimensional subspaces of the full $3d$ space; the former are called \emph{fractons}, while the latter are called \emph{lineons} or \emph{planons} depending on their mobility. 
The restricted mobility of quasiparticles in gapped fracton phases causes a ground-state degeneracy that is sensitive to the geometry of the system and often exponentially grows with increasing of the system size, but the degeneracy is still topologically stable in the sense that it cannot be split by local perturbations.

Since both geometry and topology essentially come into play, the fracton topological phases fall outside of the effective description in terms of topological quantum field theory commonly used for conventional topologically ordered phases. 
The fracton phases rather require some lattice description and in fact many key properties of gapped fracton phases have been established upon the construction of exactly solvable lattice models, which often consist of local commuting projectors. 
There have been several proposed schemes to obtain such lattice models, including the construction from coupled layers of $2d$ topological phases \cite{HMa17, Vijay17a, Vijay17b, Slagle17a, Prem19a, Shirley19d}, spin chains \cite{Halasz17}, Majorana fermions \cite{THHsieh17, YYou18b, YYou18c}, string-membrane-net condensation \cite{Slagle19}, and gauging of associated symmetry-protected topological phases \cite{Vijay16, Williamson16, Williamson18, YYou18a, Shirley19c}. 
Especially to realize fracton topological phases in experiment, the construction from constituents naturally appearing in materials will be much desired. 

In this paper, we propose a coupled-layer construction of fracton topological phases, which differs from those developed previously \cite{HMa17, Vijay17a, Prem19a, Shirley19d}; 
instead of stacking layers of $2d$ topological orders in all three orthogonal directions of the $3d$ space, our construction requires a stack of $2d$ topological orders only in one direction. 
With appropriate couplings to implement anyon condensation between layers, the corresponding models undergo phase transitions from decoupled $2d$ topological phases to fracton topological phases. 
The resulting fracton phases have quasiparticles with spatially anisotropic mobility, in contrast to the ``isotropic'' layer constructions \cite{HMa17, Vijay17a, Prem19a, Shirley19d} which yield the X-cube model \cite{Vijay16} or their relatives with the same mobility of quasiparticles in all three directions. 
While our models lack fractons as strictly immobile excitations, the models exhibit lineons constrained in one-dimensional subspaces, whose dipoles behave as planons mobile in two-dimensional subspaces 
\footnote{As a remark, while our models do not possess ``fractons'' as strictly immobile excitations, we abuse ``fracton models'' or ``fracton topological orders'' to emphasize that the corresponding models are still distinguished from the conventional topological orders or their decoulpled stacks.}.
In the special case of stacked toric codes on the square lattice, the models in the strong-coupling limit turn out to be described by exactly solvable models proposed by Shirley, Slagle, and Chen in Ref.~\cite{Shirley18a}. 
As the models obtained by our construction take simpler forms than the previous models, they may help us to seek experimental realizations of fracton topological orders, e.g., in $2d$ spin-liquid candidate materials with ``bad'' two-dimensionality. 

The rest of paper is organized as follows: 
In Sec.~\ref{sec:AnyonCondensTC}, we review anyon condensations and their model realizations in the $2d$ toric code \cite{Kitaev03}. 
In Sec.~\ref{sec:FractonFromCTC}, we present the simplest model consisting of stacked layers of the $2d$ toric codes on the square lattice, whose quasiparticle excitations in a fracton phase are analyzed in the strong-coupling limit or understood from the perspective of anyon condensation. 
In Sec.~\ref{sec:Generalization}, we propose generalizations of the construction for stacked layers of other $2d$ lattice models. 
We conclude our paper with several discussions in Sec.~\ref{sec:Conclusion}. 

\section{Anyon condensation in the toric code}
\label{sec:AnyonCondensTC}

Prior to addressing the construction of anisotropic fracton models from coupled layers of $2d$ topological orders, we first consider anyon condensation transitions in a single layer or a bilayer of the $2d$ toric code. 
This will be helpful on later interpreting the anisotropic mobility of quasiparticles in fracton phases as a result of anyon condensation and will also give us insights on how to construct the corresponding lattice Hamiltonians.

\subsection{$2d$ toric code}
\label{sec:ToricCode}

\begin{figure}
\includegraphics[clip,width=0.45\textwidth]{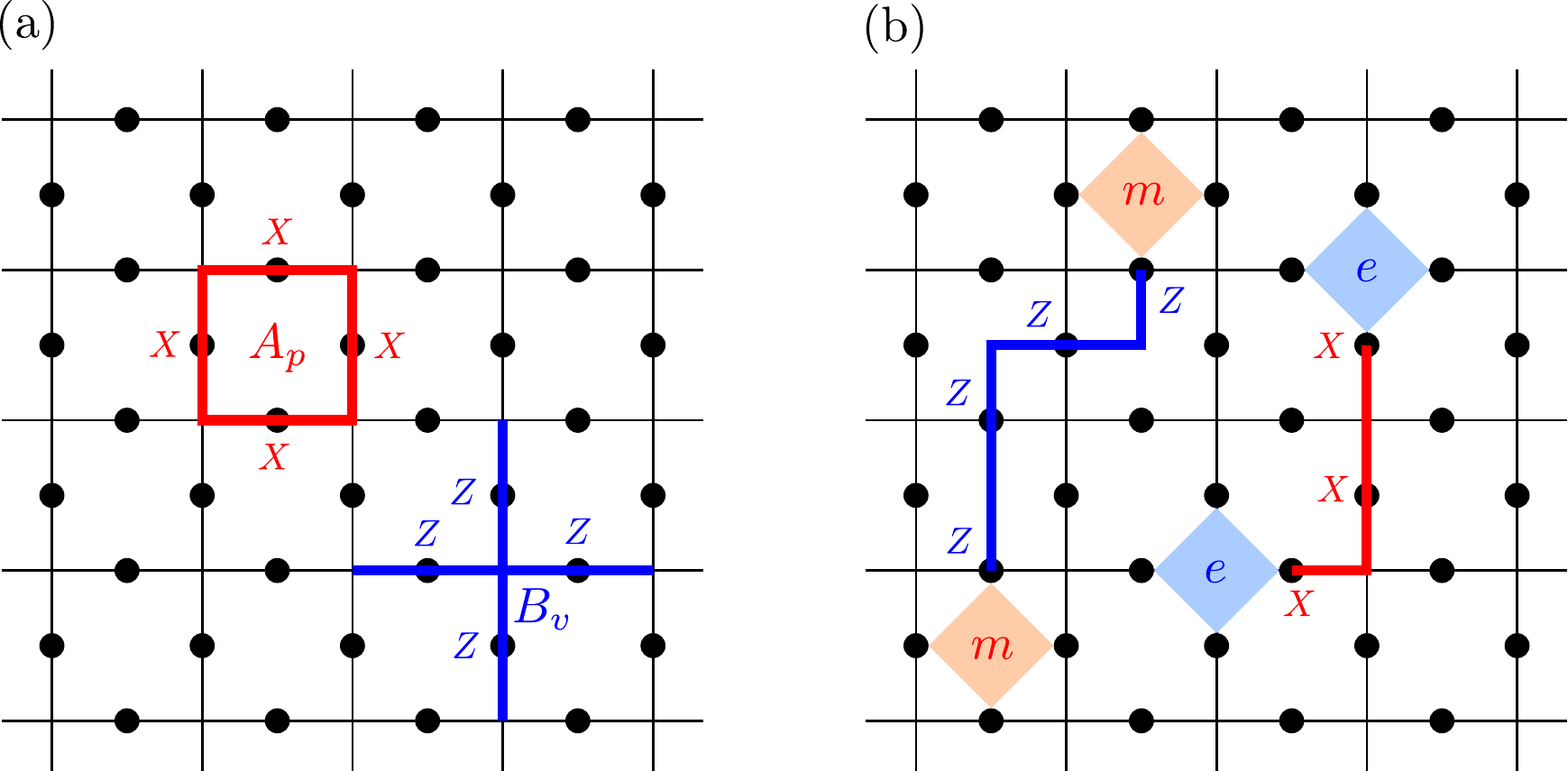}
\caption{(a) Hamiltonian for the $2d$ toric code on the square lattice. 
The black dot represents a qubit on a link.
The operator $A_p$ is the product of Pauli $X$ operators acting on four links of the plaquette $p$, while $B_v$ is the product of $Z$'s acting on four links of the vertex $v$.
(b) Deconfined quasiparticle excitations from the ground state of the toric code. 
The $e$ excitations are created on vertices by a string of $X$'s, while the $m$ excitations are created on plaquettes by a string of $Z$'s.}
\label{fig:2dTC}
\end{figure}

Let us first review properties of the $2d$ toric code on the square lattice \cite{Kitaev03}. 
We consider the system of qubits put on each link of the square lattice. 
The Hamiltonian is given by 
\begin{align} \label{eq:HamTC}
H_\textrm{TC} = -J_p \sum_p A_p -J_v \sum_v B_v, 
\end{align}
where the operators $A_p$ and $B_v$ are defined on each plaquette ($p$) or vertex ($v$) on the square lattice by 
\begin{align}
A_p &= \prod_{l \in p} X_l, \\
B_v &= \prod_{l \in v} Z_l. 
\end{align}
Here, we have defined $X_l$ and $Z_l$ as the Pauli operators acting on a qubit on the link $l$, and the products are taken over four links forming a plaquette $p$ or vertex $v$. 
The Hamiltonian is pictorially given in Fig.~\ref{fig:2dTC}~(a). 
Since the operators $A_p$ and $B_v$ satisfy $A_p^2 = B_v^2 =1$ and $[A_p, A_{p'}] = [B_v, B_{v'}] = [A_p, B_v]=0$, the ground state is obtained as a simultaneous eigenstate of $A_p$ and $B_v$ whose eigenvalues are all $+1$. 
However, not all operators can span the Hilbert space when the system is placed on a torus, due to the constraints $\prod_p A_p =1$ and $\prod_v B_p =1$, which result in the $2^2$-fold degeneracy of the ground state. 
This ground-state degeneracy is topological in the sense that it cannot be lifted by any local perturbations; as the ground-state manifold is spanned by nonlocal string operators of $X_l$ or $Z_l$ winding noncontractible cycles of the torus, splitting of the degeneracy by local perturbations is exponentially suppressed with increasing the system size.

There are two types of excitations from the ground state.
As depicted in Fig.~\ref{fig:2dTC}~(b), acting an open string of $X_l$ creates a pair of excitations at the ends of the string, where the eigenvalues of $B_v$ are flipped to be $-1$. 
Similarly, an open string of $Z_l$ creates a pair of excitations at the ends of the string, where the eigenvalues of $A_p$ are flipped. 
These excitations are deconfined since the excitation energy remains constant regardless of the length of the string. 
We may call the excitations created by $X_l$ electric charges $e$, as they live on the vertices, while those created by $Z_l$ magnetic charges $m$, as they live on the plaquettes. 
Writing their bound object as $\psi \equiv em$ and the vacuum as $\id$, the toric code has the four quasiparticles:
\begin{align}
\textrm{TC}: \ \{ \id, e, m, \psi \}
\end{align}
The same species of quasiparticles fuse to the vacuum, $e \times e = m \times m = \psi \times \psi = \id$. 
While $e$ and $m$ are bosons on their own, they have the nontrivial mutual statistics of $\pi$ since two string operators of $X_l$ and $Z_l$ intersect once and pick up a phase $-1$ when $e$ ($m$) goes around $m$ ($e$).
As a result, their bound object $\psi$ behaves as a fermion. 
They then have the fusion rules $e \times m = \psi$, $m \times \psi = e$, and $e \times \psi = m$. 
This is known as the $Z_2$ topological order.

\subsection{Condensation in a single layer}

We here review phase transitions induced by anyon condensation in a single layer of the toric code. 
The concept of anyon condensation was introduced by Bais and Slingerland \cite{Bais09} in order to discuss phase transitions between $2d$ topological orders driven by the condensation of bosonic quasiparticles, and several systematic approaches to identifying the topological orders in condensed phases have been developed \cite{Eliens14, LKong14, Neupert16}; see also a recent review \cite{Burnell18}. 
In the case of the toric code, there are two types of bosonic quasiparticle, $e$ and $m$, either of which can be condensed. 
In the $e$ condensate, remaining quasiparticles $m$ and $\psi$ both have the nontrivial mutual statistics of $\pi$ with respect to $e$ and thus are confined; the resulting state has a trivial topological order. 
Similarly, the condensation of $m$ also leads to a trivial topological order.

Let us consider microscopic models exemplifying these anyon condensation transitions. 
As reviewed in Sec.~\ref{sec:ToricCode}, $e$ excitations are created by acting the Pauli operator $X_l$'s on the ground state of the toric-code Hamiltonian \eqref{eq:HamTC}, while $m$ excitations are created by acting $Z_l$'s. 
Hence, the condensation of $e$ or $m$ will be simply induced by applying magnetic fields to the toric code: 
\begin{align}
H^e &= H_\textrm{TC} -h_X \sum_l X_l, \\
H^m &= H_\textrm{TC} -h_Z \sum_l Z_l. 
\end{align}
We expect that the magnetic field $h_X$ induces an $e$-condensation transition while $h_Z$ induces an $m$-condensation transition. 
Obviously, the ground state becomes a fully polarized state with a trivial topological order in the limit of large $h_X$ or $h_Z$, and thus there will be a phase transition between the $Z_2$ and trivial topological orders for some $h_X$ or $h_Z$. 
In fact, these models have been studied previously \cite{Trebst07, JVidal09a, Tupitsyn10, Dusuel11, FWu12}; the corresponding phase transitions are conjectured to be in the (2+1)-$d$ Ising${}^*$ universality class---the Wilson-Fisher fixed point of a real scalar field theory coupled with a $Z_2$ gauge field \cite{Schuler16}. 

\subsection{Condensation between two layers}
\label{sec:CondensTwoLayers}

We then focus on two layers of the $2d$ toric codes. 
When the two layers are decoupled, the ground state has a topological order with 16 quasiparticles given by 
\begin{align} \label{eq:DoubledTCQPs}
\textrm{TC}_1 \otimes \textrm{TC}_2: \ \{ \id, e, m, \psi \}_1 \otimes \{ \id, e, m, \psi \}_2, 
\end{align}
where the subscripts $1$ and $2$ refer to the two layers and the quasiparticles are simply given by tensor products of those from each layer. 
We then consider phase transitions induced by the condensation of bound pairs of quasiparticles between the two layers. 
For examples, $e_1 e_2$ is a boson and thus can be condensed. 
In the condensate, all quasiparticles obeying nontrivial mutual statistics with $e_1 e_2$ are confined. 
We are thus left with a topological order characterized by the following quasiparticles: 
\begin{align}
\tTC: \ \{ \id_1 \id_2, e_1 {\bm 1}_2, m_1 m_2, \psi_1 m_2 \} \equiv \{ \tid, \te, \tm, \tpsi \}. 
\end{align}
Here, quasiparticles that are transformed to each other by the fusion with $e_1 e_2$ are identified. 
The resulting topological order has four quasiparticles, and their statistics exactly matches with that for a single layer of the toric code; we thus call it as $\tTC$ and label the quasiparticles by symbols with a tilde. 

\begin{figure}
\includegraphics[clip,width=0.45\textwidth]{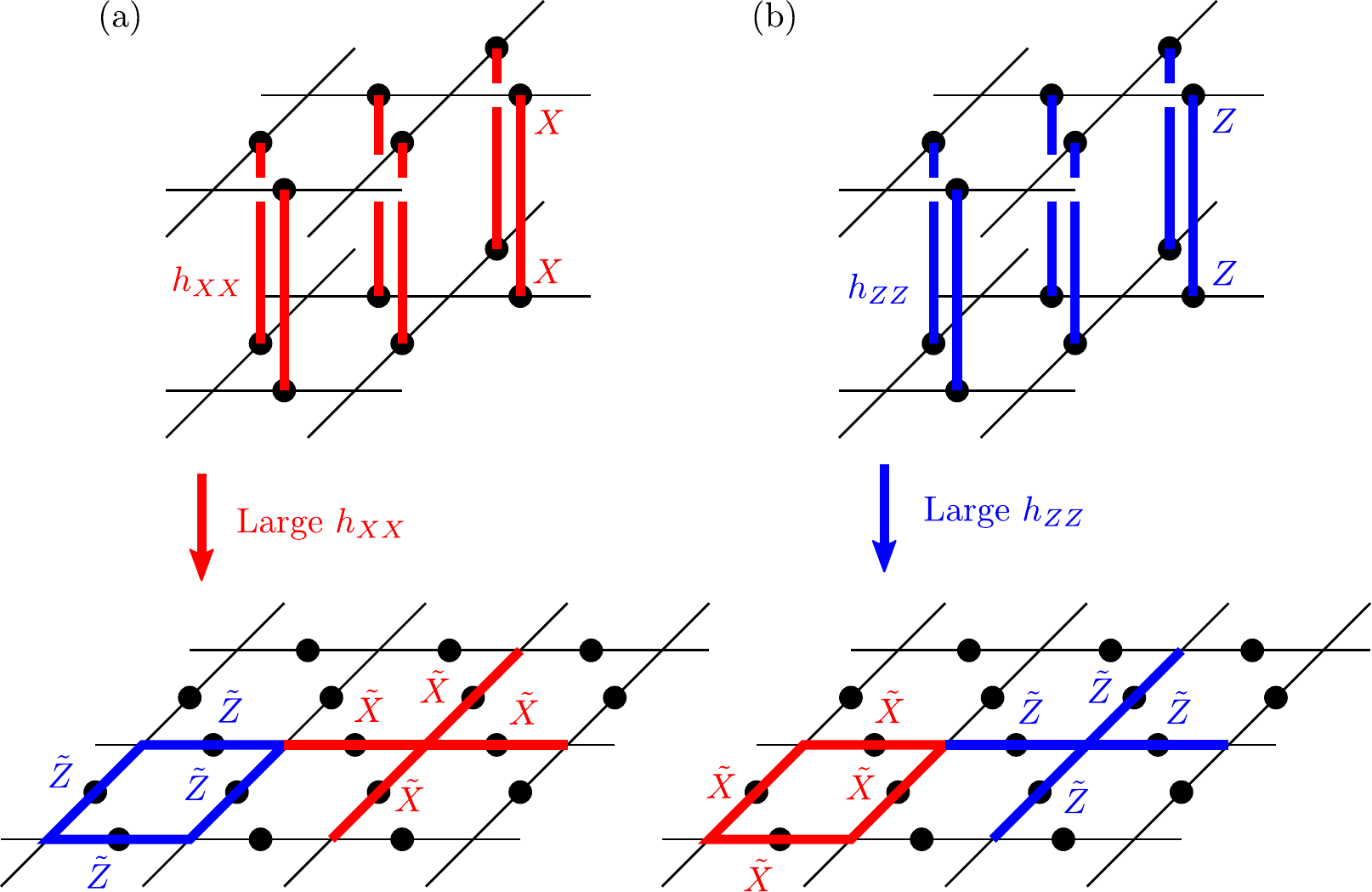}
\caption{Coupling between two toric codes and the effective Hamiltonian in the strong-coupling limit for (a) the $e_1 e_2$ condensation and (b) the $m_1 m_2$ condensation.}
\label{fig:BilayerTC}
\end{figure}

We want to implement this phenomenology of the topological phase transition from $\textrm{TC}_1 \otimes \textrm{TC}_2$ to $\tTC$ in a microscopic model. 
Taking a bilayer of the toric code, a bound pair of $e$'s from each layer may be created by the action of $X_{l,1} X_{l,2}$ where the Pauli operator $X_{l,n}$ acts on a qubit on the link $l$ of the layer $n$. 
We thus consider the Hamiltonian, 
\begin{align} \label{eq:HamTCee}
H^{ee} = H_{\textrm{TC},1} + H_{\textrm{TC},2} -h_{XX} \sum_l X_{l,1} X_{l,2},
\end{align}
where
\begin{align} \label{eq:ToricCodeOnN}
H_{\textrm{TC},n} = -J_p \sum_p \prod_{l \in p} X_{l,n} -J_v \sum_v \prod_{l \in v} Z_{l,n}. 
\end{align}
Again, the labels $p$ and $v$ refer to the plaquettes and vertices of the square lattice, respectively, as shown in Fig.~\ref{fig:2dTC}~(a). 
The structure of the coupling between layers is schematically given in Fig.~\ref{fig:BilayerTC}~(a).
In order to see that this Hamiltonian exhibits the expected phase transition, we now consider the strong-coupling limit $h_{XX} \to \infty$. 
In this limit, the two qubits on the same link $l$ are in either of two states, $\left| + \right>_{l,1} \left| + \right>_{l,2}$ or $\left| - \right>_{l,1} \left| - \right>_{l,2}$, in the basis diagonalizing $X_{l,n}$, i.e. $X_{l,n} \left| \pm \right>_{l,n} = \pm \left| \pm \right>_{l,n}$. 
Let us denote by $\left| \tpm \right>_l \equiv \left| \pm \right>_{l,1} \left| \pm \right>_{l,2}$ two basis states of an effective qubit on the link $l$. 
We then perform degenerate perturbation theory in the Hilbert space of effective qubits by treating $H_{\textrm{TC},n}$ as perturbations. 
In doing so, we define the Pauli operators $\tX_l$ and $\tZ_l$ acting on the effective qubits and choose $\tZ_l$ to be diagonal, i.e. $\tZ_l \left| \tpm \right>_l = \pm \left| \tpm \right>_l$ and $\tX_l \left| \tpm \right>_l = \left| \tmp \right>_l$.
Projecting the original Pauli operators into the subspace spanned by effective qubits, we find $X_{l,1} = X_{l,2} = \tZ_l$ and $Z_{l,1} Z_{l,2} = \tX_l$. 
Up to the second order in perturbation, we arrive at the effective Hamiltonian, 
\begin{align}
H_\textrm{eff}^{ee} = -2J_p \sum_p \prod_{l \in p} \tZ_l -\frac{J_v^2}{8h_{XX}} \sum_v \prod_{l \in v} \tX_l.
\end{align}
This Hamiltonian is equivalent to that for the $2d$ toric code up to a unitary transformation. 
Since the ground state is expected to have the $Z_2$ topological order for a large enough $h_{XX}$, we conclude that the Hamiltonian \eqref{eq:HamTCee} describes a phase transition from $\textrm{TC}_1 \otimes \textrm{TC}_2$ to $\tTC$ induced by the $e_1 e_2$ condensation.

For the $m_1 m_2$ condensation in $\textrm{TC}_1 \otimes \textrm{TC}_2$, the quasiparticle content in the condensate is given by 
\begin{align}
\tTC: \ \{ {\bm 1}_1 {\bm 1}_2, e_1 e_2, m_1 {\bm 1}_2, \psi_1 e_2 \} \equiv \{ \tid, \te, \tm, \tpsi \}, 
\end{align}
and is the same as that for a single layer of the toric code. 
As the Pauli operator $Z_{l,n}$ creates $m$ excitations on each layer, the corresponding Hamiltonian may be given by 
\begin{align} \label{eq:HamTCmm}
H^{mm} = H_{\textrm{TC},1} +H_{\textrm{TC},2} -h_{ZZ} \sum_l Z_{l,1} Z_{l,2}
\end{align}
as shown in Fig.~\ref{fig:BilayerTC}~(b). 
In the strong-coupling limit $h_{ZZ} \to \infty$, we can introduce the basis states for effective qubits as $\left| \tpm \right>_l \equiv \left| \pm \right>_{l,1} \left| \pm \right>_{l,2}$ where we have chosen the original basis to be $Z_{l,n} \left| \pm \right>_{l,n} = \pm \left| \pm \right>_{l,n}$. 
By performing degenerate perturbation theory, we find the effective Hamiltonian
\begin{align}
H^{mm}_\textrm{eff} = -\frac{J_p^2}{8h_{ZZ}} \sum_p \prod_{l \in p} \tX_l -2J_v \sum_v \prod_{l \in v} \tZ_l, 
\end{align}
where we have defined the Pauli operators acting on the effective qubits in such a way that $\tZ_l \left| \tpm \right>_l = \pm \left| \tpm \right>_l$ and $\tX_l \left| \tpm \right>_l = \left| \tmp \right>_l$.
This is again a single layer of the $2d$ toric code, although this result is not surprising from the $e$-$m$ duality in the $2d$ toric code.
Thus, the Hamiltonian \eqref{eq:HamTCmm} will describe a topological phase transition induced by the $m_1 m_2$ condensation.

Before proceeding, we make a remark about the nature of the transitions. 
While it is not so obvious, the transition from $\textrm{TC}_1 \otimes \textrm{TC}_2$ to $\tTC$ induced by the $e_1e_2$ or $m_1 m_2$ condensation is expected to be in the Ising${}^*$ universality class as in the single-layer case.
This is because the transition in the bilayer can be viewed as a single-layer transition from the $Z_2$ to trivial topological order from its quasiparticle content; 
we can rewrite the quasiparticle content of $\textrm{TC}_1 \otimes \textrm{TC}_2$ as 
\begin{align}
\{ \id_1 \id_2, e_1 e_2, \id_1 m_2, e_1 \psi_2 \} \times \{ \id_1 \id_2, e_1 \id_2, m_1 m_2, \psi_1 m_2 \}, 
\end{align}
where the multiplication should be operated in the sense of fusion. 
Since the two sets of quasiparticles both represent the $Z_2$ topological order while the statistics of quasiparticles are mutually trivial between the two sets, we can regard them as two decoupled layers of the $Z_2$ topological orders. 
The $e_1 e_2$ or $m_1 m_2$ condensation is simply viewed as the single $\te$ or $\tm$ condensation in one layer with leaving another layer intact. 
Therefore, the associate transition is naturally expected to be of the Ising${}^*$ type. 

\section{Fracton model from coupled toric codes}
\label{sec:FractonFromCTC}

We present a model that may have an anyon condensation transition from decoupled layers of the $2d$ toric codes to a nontrivial fracton topological order. 
In the limit of strong coupling between layers, we can write down an effective Hamiltonian that is exactly solvable. 
In fact, the resulting model has been proposed in Ref.~\cite{Shirley18a} and possesses fractionalized quasiparticles with spatially anisotropic mobility, which is yet different from that of the stacked toric codes. 
We argue that the anisotropic mobility of quasiparticles can be naturally explained in terms of anyon condensation induced by coupling between the toric codes.

\subsection{Model}

\begin{figure}
\includegraphics[clip,width=0.45\textwidth]{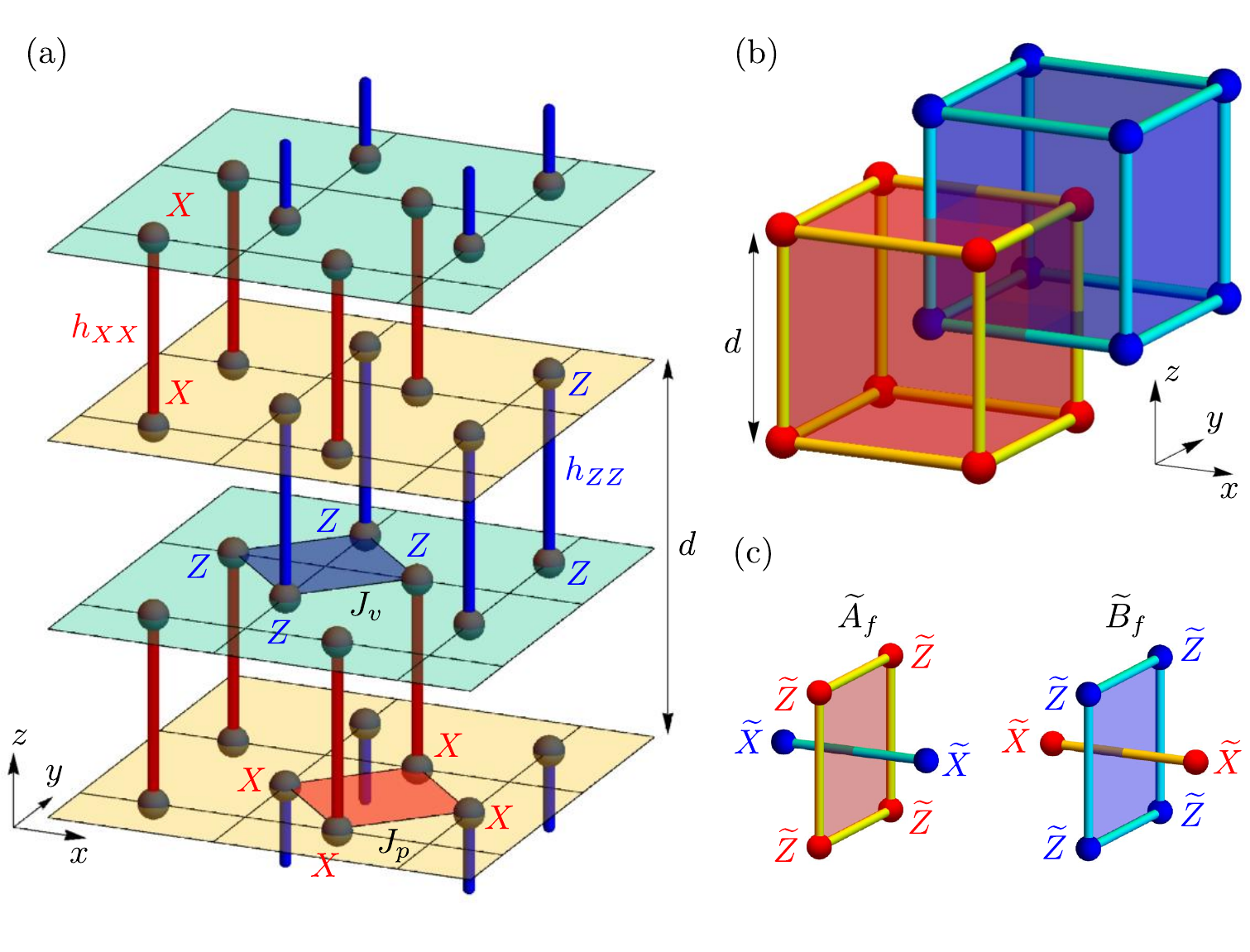}
\caption{(a) Coupled-layer model from the $2d$ toric codes. 
The red (blue) bonds represent $h_{XX}$ ($h_{ZZ}$) couplings between adjacent layers. 
The red (blue) plaquette represents the interaction $J_p$ ($J_v$) of the toric code. 
Since the couplings take a staggered structure, we denote by $d$ an enlarged unit distance along the $z$ axis. 
(b) The effective Hamiltonian \eqref{eq:AnisotropicFracton} in the strong-coupling limit is defined on a bcc lattice whose $A$ ($B$) sublattice consists of effective qubits on the $h_{XX}$ ($
h_{ZZ}$) bonds. 
(c) Local terms are composed of four $\tilde{Z}$ operators acting on the corners of a $yz$ face of a cube and two $\tilde{X}$ operators acting on the ends of $x$ bonds normals to the face.}
\label{fig:FractonFromTC}
\end{figure}

We consider layers of the $2d$ toric code lying in the $xy$ plane and stacked along the $z$ axis, which are given by the Hamiltonian,
\begin{align} \label{eq:StackedTC}
H_1 = \sum_{n} H_{\textrm{TC},n}
\end{align}
where $H_{\textrm{TC},n}$ is the Hamiltonian for the toric code on the $n$-th layer and is defined in Eq.~\eqref{eq:ToricCodeOnN}.
We then consider coupling between layers that takes a staggered structure as follows: 
Two qubits from the $2m$-th and ($2m+1$)-th layers are coupled via $X_{l,2m} X_{l,2m+1}$ on the link $l$ aligned with the $x$ axis, whereas two qubits from the ($2m-1$)-th and $2m$-th layers are coupled via $Z_{l,2m-1} Z_{l,2m}$ on the link $l$ aligned with the $y$ axis. 
The corresponding Hamiltonian is given by 
\begin{align}
H_0 &= \sum_m \biggl[ -h_{XX} \sum_{l \parallel x} X_{l,2m} X_{l,2m+1} \nonumber \\
&-h_{ZZ} \sum_{l \parallel y} Z_{l,2m-1} Z_{l,2m} \biggr],
\end{align}
which is schematically shown in Fig.~\ref{fig:FractonFromTC}~(a).
The full Hamiltonian for the coupled toric codes is then given by
\begin{align} \label{eq:HamCTC}
H_\textrm{CTC} = H_0 +H_1. 
\end{align}
The ground state obviously has the topological order of decoupled $2d$ toric codes for $h_{XX}=h_{ZZ}=0$, which is characterized by quasiparticles of the $Z_2$ topological order deconfined only within each layer and the ground-state degeneracy $4^{L_z}$ on a three-torus with $L_z$ layers. 
The model is no longer exactly solvable for a general choice of the parameters. 
However, an effective Hamiltonian obtained in the strong-coupling limit $h_{XX}, h_{ZZ} \to \infty$ is exactly solvable as we will see below.

\subsection{Strong-coupling limit: Anisotropic fracton model}
\label{sec:StrongCouplingAF}

In the spirit of Sec.~\ref{sec:CondensTwoLayers}, we here derive an effective Hamiltonian in the strong-coupling limit $h_{XX},h_{ZZ} \to \infty$. 
In the limit $h_{XX} \to \infty$, two qubits on the link $l$ parallel to the $x$ axis form either of the two states  between the $2m$-th and ($2m+1$)-th layers,
\begin{align}
\begin{split}
\left| \tplus \right>_{l,2m+1/2} &\equiv \left| + \right>_{l,2m} \left| + \right>_{l,2m+1}, \\
\left| \tminus \right>_{l,2m+1/2} &\equiv \left| - \right>_{l,2m} \left| - \right>_{l,2m+1}, 
\end{split}
\end{align}
in the basis where $X_{l,n}$ is diagonalized: $X_{l,n} \left| \pm \right>_{l,n} = \pm \left| \pm \right>_{l,n}$. 
In the limit $h_{ZZ} \to \infty$, two qubits on the link $l$ parallel to the $y$ axis form either of the two states  between the ($2m-1$)-th and $2m$-th layers,
\begin{align}
\left| \tplus \right>_{l,2m-1/2} &\equiv \left| + \right>_{l,2m-1} \left| + \right>_{l,2m}, \\
\left| \tminus \right>_{l,2m-1/2} &\equiv \left| - \right>_{l,2m-1} \left| - \right>_{l,2m}, 
\end{align}
in the basis where $Z_{l,n}$ is diagonalized: $Z_{l,n} \left| \pm \right>_{l,n} = \pm \left| \pm \right>_{l,n}$. 
We now treat the toric-code Hamiltonian $H_1$ in Eq.~\eqref{eq:StackedTC} as a perturbation and perform degenerate perturbation theory to obtain an effective Hamiltonian acting on the Hilbert space of effective qubits $\left| \tpm \right>$. 
By squashing the $XX$ and $ZZ$ bonds between layers to points, these effective qubits can be viewed to live on a bcc lattice as shown in Fig.~\ref{fig:FractonFromTC}~(b); 
let us denote by the $A$ sublattice a cubic lattice composed of effective qubits on the ($2m+1/2$)-th layers, which used to be defined on the $XX$ bonds of the coupled toric codes, whereas by the $B$ sublattice another cubic lattice composed of effective qubits on the ($2m-1/2$)-th layers, which used to be defined on the $ZZ$ bonds. 
We then introduce the Pauli operators $\tX^{A(B)}_s$ and $\tZ^{A(B)}_s$ acting on the effective qubit at the site $s$ on the $A$ ($B$) sublattice of the bcc lattice. 
After the projection onto the Hilbert space of effective qubits, the original Pauli operators $X_{l,n}$ and $Z_{l,n}$ can be represented as 
\begin{align}
\begin{split}
X_{l,2m} = X_{l,2m+1} &= \tZ^A_s, \\
Z_{l,2m} Z_{l,2m+1} &= \tX^A_s
\end{split}
\end{align}
for $l \parallel x$, while 
\begin{align}
\begin{split}
X_{l,2m-1} X_{l,2m} &= \tX^B_s, \\
Z_{l,2m-1} = Z_{l,2m} &= \tZ^B_s
\end{split}
\end{align}
for $l \parallel y$. 

Up to the second order in $H_1$, we find the effective Hamiltonian acting on effective qubits,
\begin{align} \label{eq:AnisotropicFracton}
H^\textrm{eff}_\textrm{CTC} = -\frac{J_p^2}{4h_{ZZ}} \sum_{\substack{f \in A \\ f \parallel yz}} \tA_f -\frac{J_v^2}{4h_{XX}} \sum_{\substack{f \in B \\ f \parallel yz}} \tB_f, 
\end{align}
where $f$ denotes a face of a cube that belongs to either $A$ or $B$ sublattice and \emph{is parallel to the $yz$ plane}, and $\tA_f$ and $\tB_f$ are local operators defined by
\begin{align}
\begin{split}
\tA_f &= \prod_{s \in f} \tZ^A_s \prod_{s' \perp f} \tX^B_{s'}, \\
\tB_f &= \prod_{s \in f} \tZ^B_s \prod_{s' \perp f} \tX^A_{s'}.
\end{split}
\end{align}
As seen from Fig.~\ref{fig:FractonFromTC}~(c), the operator $\tA_f$ is a product of four $\tZ$ operators on the $A$ sublattice at the corners of the face $f$ and two $\tX$ operators on the $B$ sublattice at the ends of a bond normal to the face $f$, and similarly for $\tB_f$ by interchanging the $A$ and $B$ sublattices. 
The same Hamiltonian as Eq.~\eqref{eq:AnisotropicFracton} has been previously introduced in Ref.~\cite{Shirley18a} for an anisotropic fracton model. 
While basic properties of the Hamiltonian have already been discussed in the same reference, we review below those properties in detail for completeness.

\subsubsection{Ground-state degeneracy}
\label{sec:GSD}

As any two operators from $\tA_f$ and $\tB_f$ commute, the effective Hamiltonian \eqref{eq:AnisotropicFracton} is exactly solvable. 
Since $\tA_f^2 = \tB_f^2=1$, the ground state is given by a simultaneous eigenstate of $\tA_f$ and $\tB_f$ whose eigenvalues are all $+1$. 
Similarly to the toric code, there can be constraints that make certain products of the operators $\tA_f$ and $\tB_f$ to be the identity, leaving the ground-state degeneracy. 
Let us consider the system put on a three-torus with the linear sizes $L_x \times L_y \times L_z$ such that there are $2L_x L_y L_z$ effective qubits.
We may first multiply $\tA_f$ along the $x$ axis to cancel the $\tX^B$ operators. 
Residual $\tZ^A$ operators form a ``tube'' along the $x$ axis and can be multiplied along either the $y$ or $z$ axis to be the identity. 
We thus find the following constraints,
\begin{align}
\prod_{f \parallel xy} \tA_f = \prod_{f \parallel xz} \tA_f =1, 
\end{align}
meaning that $\tA_f$'s multiplied over the $xy$ or $xz$ plane become the identity.
Similarly, we also have
\begin{align}
\prod_{f \parallel xy} \tB_f = \prod_{f \parallel xz} \tB_f =1.
\end{align}
There are in total $2(L_y+L_z-1)$ independent conditions, which result in the subextensive ground-state degeneracy
\begin{align}
\textrm{GSD} = 2^{2(L_y+L_z-1)}.
\end{align}

\begin{figure}
\includegraphics[clip,width=0.35\textwidth]{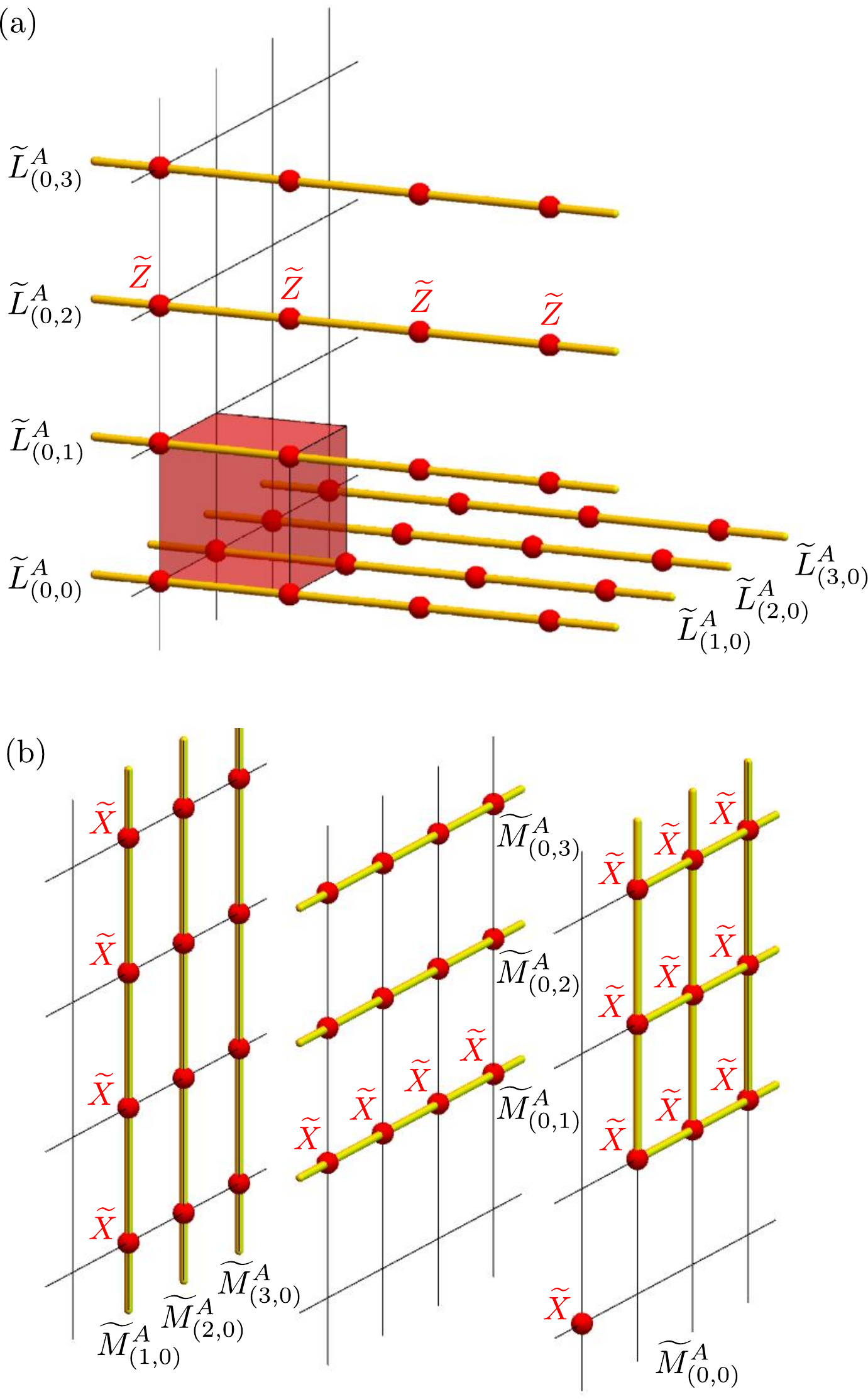}
\caption{(a) Line-like operators $\tL^A_{(y,z)}$ on a torus of the linear sizes $L_x \times 4 \times 4$. 
(b) Membrane-like operators $\tM^A_{(y,z)}$. }
\label{fig:String}
\end{figure}

The subextensive degeneracy is a signal of fracton topological order. 
However, we have to make sure that this degeneracy cannot be split by local perturbations. 
In order to see this, we here show that logical operators spanning the ground-state manifold have nonlocal supports that grow with increasing the systems size. 
This implies that splitting of degenerate ground states by local perturbations vanishes in the thermodynamic limit. 
There are line-like logical operators composed of the Pauli $\tZ$ operators on the $A$ sublattice along the $x$ axis, 
\begin{align}
\tL^A_{(y,z)} = \prod_{x=0}^{L_x-1} \tZ^A_{(x,y,z)}, 
\end{align}
where $s=(x,y,z)$ denotes the coordinates on the cubic lattice.
These operators commute with all terms in the Hamiltonian \eqref{eq:AnisotropicFracton}. 
If a product of the operators $\tL^A_{(y,z)}$ forms a quadrangular prism whose base is a rectangle on the $yz$ plane, it can be written in terms of a product of $\tA_f$ and trivially acts on the ground state. 
Thus, there are only $L_y-L_z-1$ independent line-like operators nontrivially acting on the ground state. 
For our convenience, we make the following choice for the coordinates $(y,z)$ of such line-like operators,
\begin{align}
\{ (y,z) \} &= \{ (0,0) \} \cup \{ (y,0) | y=1, \cdots, L_y-1 \} \nonumber \\
&\cup \{ (0,z) | z=1, \cdots, L_z-1 \}
\end{align}
as shown in Fig.~\ref{fig:String}~(a). 
In addition, there are also ``membrane-like'' logical operators composed of the Pauli $\tX$ operators on the $A$ sublattice in a $yz$ plane, 
\begin{align}
\begin{split}
\tM^A_{(y,0)} &= \prod_{z=0}^{L_z-1} \tX^A_{(x,y,z)}, \hspace{10pt} y=1, \cdots, L_y-1 \\
\tM^A_{(0,z)} &= \prod_{y=0}^{L_y-1} \tX^A_{(x,y,z)}, \hspace{10pt} z=1, \cdots, L_z-1 \\
\tM^A_{(0,0)} &= \tX^A_{(x,0,0)} \prod_{y=1}^{L_y-1} \prod_{z=1}^{L_z-1} \tX^A_{(x,y,z)}, 
\end{split}
\end{align}
as shown in Fig.~\ref{fig:String}~(b). 
Here, the choice of $x$ is arbitrary since $\tM^A_{(y,z)}$ can be shifted along the $x$ axis by multiplying operators $\tB_f$. 
Again, they commute with all terms in the Hamiltonian and nontrvially act on the ground state. 
Upon our choice, the line-like operators $\tL^A_{(y,z)}$ and membrane-like operators $\tM^A_{(y,z)}$ anticommute for the same $(y,z)$ but commute for different $(y,z)$'s: 
\begin{align}
\tL^A_{(y,z)} \tM^A_{(y',z')} = (-1)^{\delta_{yy'} \delta_{zz'}} \tM^A_{(y',z')} \tL^A_{(y,z)}. 
\end{align}
We can similarly construct the line-like and membrane-like operators on the $B$ sublattice for which a similar algebra holds.
These $2(L_y-L_z-1)$ sets of logical operators span the $2^{2(L_y-L_z-1)}$-dimensional Hilbert space of the degenerate ground-state manifold. 
Importantly, these logical operators cannot be multiplied to form any local operators. 
This implies that a matrix element between degenerate ground states is generated by local perturbations at least at the order of $L_x$, $L_y$, or $L_z$ and is expected to vanish in the thermodynamic limit. 
This ensures a topological stability of the subextensive degeneracy and thereby a fracton topological order.

\subsubsection{Subdimensional excitations}

The subextensive ground-state degeneracy computed above is a consequence of deconfined excitations restricted in lower-dimensional subspaces of the $3d$ space. 
Since the local terms in the Hamiltonian, $\tA_f$ and $\tB_f$, have eigenvalues $+1$ in the ground state, excited states are obtained by flipping some of the eigenvalues by acting a local operator on the ground state. 
As the operators $\tA_f$ ($\tB_f$) are centered at faces of cubes that belong to the $A$ ($B$) sublattice and are parallel to the $yz$ plane, we may regard that excitations are created on these faces. 
Depending on the mobility of excitations, which we will see below, they are called ``lineons'' or ``planons'' in Ref.~\cite{Shirley18a}.

\begin{figure}
\includegraphics[clip,width=0.3\textwidth]{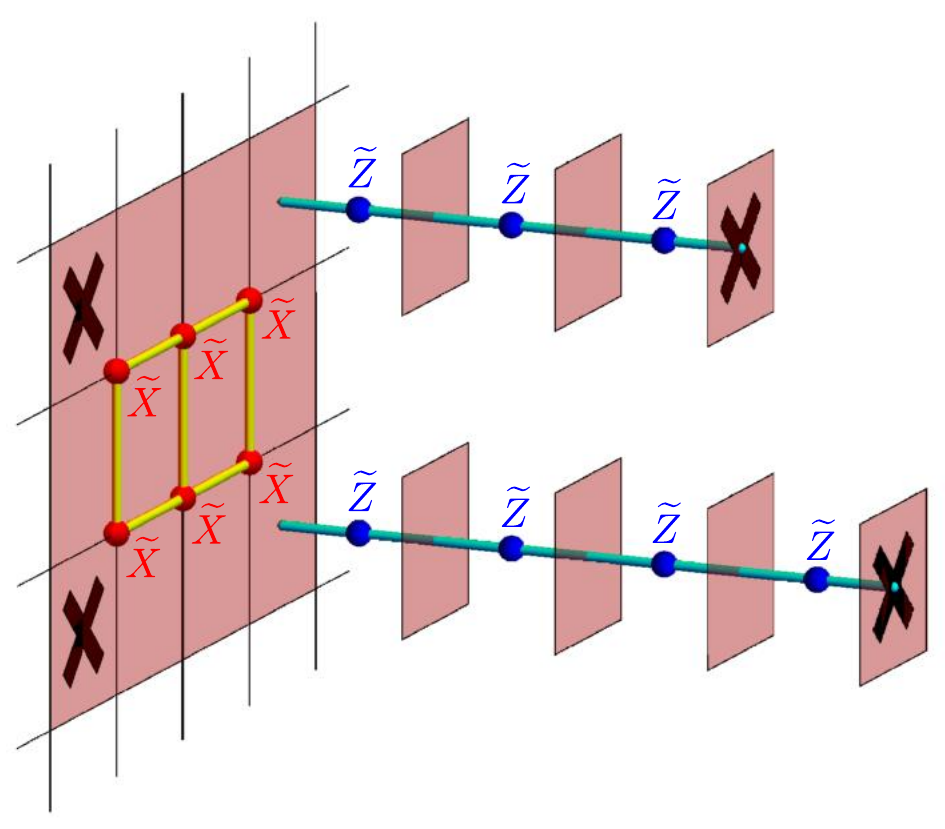}
\caption{Excitations created in the anisotropic fracton model. 
They are marked by the black crosses on faces where the eigenvalues of $\tA_f$ are flipped. 
A single excitation moves along the $x$ axis as a lineon by acting $\tZ^B$ operators, whereas a dipole of excitations separated along the $z$ axis moves within the $xy$ plane as a planon by acting $\tZ^B$ and $\tX^A$ operators.}
\label{fig:Excitations}
\end{figure}

Acting a Pauli $\tZ^B_s$ operator on the ground state, it creates excitations on two faces of the $A$ sublattice sandwiching the site $s$. 
By successively applying Pauli $\tZ^B_s$ operators, a single excitation can be transferred on a straight line along the $x$ axis. 
From its one-dimensional nature, every single excitation is called a \emph{lineon}. 
On the other hand, when two pairs of lineon excitations are created within a $xz$ plane, a dipole of excitations separated along the $z$ axis can be transferred along the $y$ axis by successively applying Pauli $\tX^A$ operators to form a rectangular membrane in the $yz$ plane; such a membrane operator by itself creates four excitations at the corners of the rectangle. 
Hence, the dipole freely moves within the $xy$ plane.
Similarly, a dipole of excitations separated along the $y$ axis can move in the $xz$ plane. 
Thus, dipoles of excitations have a $2d$ nature and are called \emph{planons}.
A way of creating such excitations is illustrated in Fig.~\ref{fig:Excitations}. 
Similarly, we can create lineon and planon excitations living on faces of the $B$ sublattice by applying Pauli $\tZ^A$ and $\tX^B$ operators.
In fact, a finite segment of the line-like operator $\tL$ creates a pair of lineon excitations along the $x$ axis, while a portion of the membrane-like operator $\tM$ creates a pair of dipole excitations in the $yz$ plane. 
When excitations are annihilated in pairs by wrapping the torus, the corresponding nonlocal operators map the ground-state manifold to itself and define logical operators $\tL$'s or $\tM$'s. 
As the excitations behave in quite different ways between the three spatial directions, the model \eqref{eq:AnisotropicFracton} is dubbed as the anisotropic fracton model.

\subsection{Anyon condensation picture}
\label{sec:AFAnyonCondens}

\begin{figure}
\includegraphics[clip,width=0.45\textwidth]{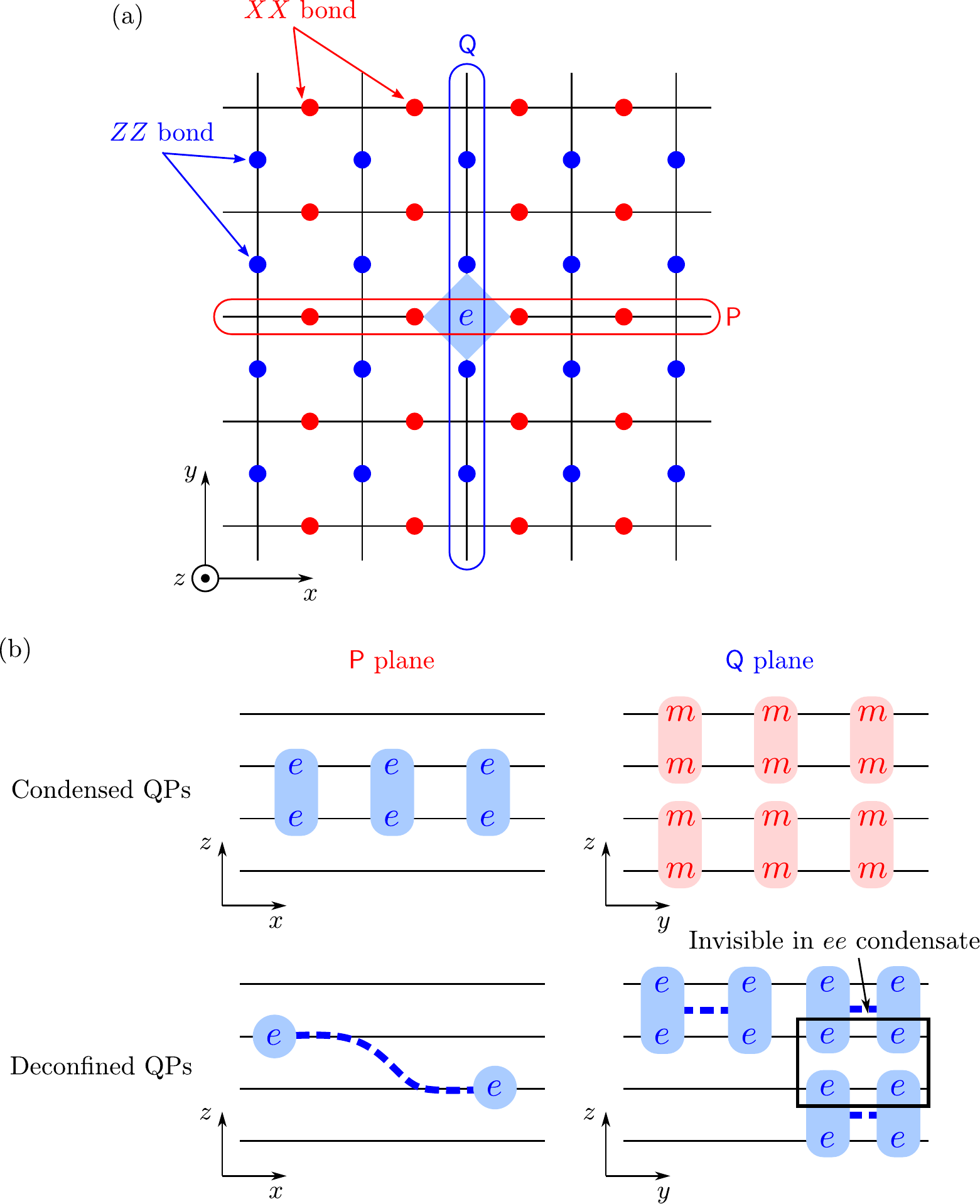}
\caption{(a) Slice of the stacked toric-code model within the $xy$ plane. 
The red (blue) circles represent the $XX$ ($ZZ$) couplings between layers, which induce the $ee$ ($mm$) condensation. 
Focusing on an $e$ excitation created on a vertex, we can consider its motion in a plane labeled by $\textsf{P}$ ($\textsf{Q}$) parallel to the $xz$ ($yz$) plane. 
(b) Condensed and deconfined quasiparticles within the $\textsf{P}$ and $\textsf{Q}$ planes.}
\label{fig:AnyonCondensAF}
\end{figure}

We can interpret the spatially anisotropic mobility of subdimensional excitations beyond the strong-coupling limit in the view of anyon condensation in the coupled toric-code model \eqref{eq:HamCTC}. 
Taking a slice of the model in the $xy$ plane, as shown in Fig.~\ref{fig:AnyonCondensAF}~(a), let us consider the mobility of an $e$ excitation created on a vertex by acting the Pauli $X$ operator.
Along the $x$ axis, the $XX$ couplings induce the $ee$ condensation between adjacent layers for a certain strength of $h_{XX}$. 
As $e$ has the trivial mutual statistics with $ee$, a single $e$ excitation can go through the $ee$ condensate, as shown in the left panel of Fig.~\ref{fig:AnyonCondensAF}~(b); this may be viewed as the origin of lineon excitations on the $B$ sublattice. 
On the other hand, the $ZZ$ couplings run along the $y$ axis and induce the $mm$ condensation between adjacent layers for a certain $h_{ZZ}$. 
Since $e$ has the nontrivial mutual statistics of $\pi$ with $mm$, $e$ excitations must form a bound pair between the adjacent layers to move in the $mm$ condensate, as shown in the right panel of Fig.~\ref{fig:AnyonCondensAF}~(b); this explains the dipole nature of excitations along the $y$ axis on the $B$ sublattice. 
When these dipoles of $e$ excitations are created successively over several layers, the excitations in internal layers become invisible due to the formation of $ee$ condensate and only $e$ excitations on the top and bottom layers are left; this establishes the dipole nature of excitations along the $z$ axis, and together with the above arguments we find planon excitations in the $xy$ or $xz$ plane. 
The same phenomenology also applies to $m$ excitations, which see the $mm$ condensate along the $x$ axis while the $ee$ condensate along the $y$ axis, leading to lineon and planon excitations on the $A$ sublattice. 

\section{Generalization}
\label{sec:Generalization}

We here generalize the construction of the anisotropic fracton model to coupled layers of other $2d$ lattice models: the Kitaev honeycomb model, the $Z_N$ toric code, and the toric code and the doubled semion model on the honeycomb lattice. 

\subsection{Kitaev-honeycomb model}
\label{sec:KitaevHoneycomb}

\begin{figure}
\includegraphics[clip,width=0.35\textwidth]{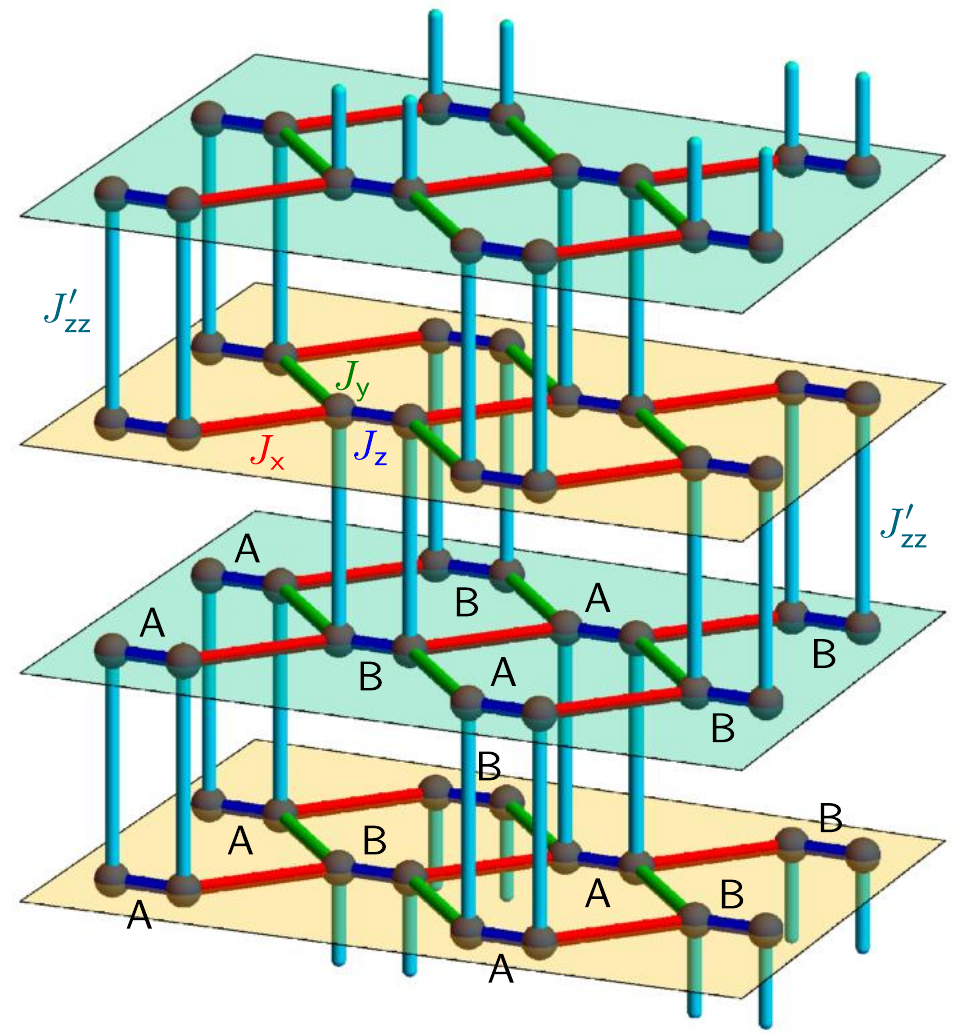}
\caption{Coupled-layer model from the Kitaev honeycomb models. 
Each layer represents the Kitaev honeycomb model with the $J_\sfx$, $J_\sfy$, and $J_\sfz$ couplings represented by the red, green, and blue bonds, respectively. 
The couplings between adjacent layers ($J'_{\sfz \sfz}$) are represented by the cyan bonds. 
The symbols $\sfA$ and $\sfB$ indicate links belonging to two different sublattices. }
\label{fig:KHLayer}
\end{figure}

Since the anisotropic fracton model \eqref{eq:AnisotropicFracton} is obtained from stacked layers of the $2d$ toric codes on the square lattice, it should be possible to realize the same model from stacked layers of the Kitaev honeycomb model \cite{Kitaev06}, as it gives the toric code in the easy-axis limits. 
The idea of constructing fracton topological order from stacked layers of the Kitaev honeycomb models has already appeared in Ref.~\cite{Slagle17a}, where stacks in two spatial directions have been used to obtain the X-cube model. 
Our construction is relatively simple as it only requires a stack in one direction. 
Let us consider the following Hamiltonian, 
\begin{align} \label{eq:StackedKitaevHoneycomb}
H_\textrm{CKH} &=  \sum_n H_\textrm{KH,n} \nonumber \\
&-J'_{\sfz \sfz} \sum_m \Biggl[ \sum_{\substack{\langle ij \rangle \in \textrm{$\sfz$-bond} \\ \langle ij \rangle \in \textrm{$\sfA$-link}}} (\sigma^\sfz_{i,2m} \sigma^\sfz_{i,2m+1} +\sigma^\sfz_{j,2m} \sigma^\sfz_{j,2m+1}) \nonumber \\
&+\sum_{\substack{\langle ij \rangle \in \textrm{$\sfz$-bond} \\ \langle ij \rangle \in \textrm{$\sfB$-link}}} (\sigma^\sfz_{i,2m-1} \sigma^\sfz_{i,2m} +\sigma^\sfz_{j,2m-1} \sigma^\sfz_{j,2m}) \Biggr], 
\end{align}
where $H_\textrm{KH,n}$ is the Kitaev honeycomb model on the $n$-th layer,
\begin{align}
H_\textrm{KH,n} = -\sum_{\alpha=\sfx,\sfy,\sfz} J_\alpha \sum_{\langle ij \rangle \in \textrm{$\alpha$-bond}} \sigma^\alpha_{i,n} \sigma^\alpha_{j,n},
\end{align}
and $\sigma^\alpha_{i,n}$ ($\alpha=\sfx,\sfy,\sfz$) are the Pauli operators at the site $i$ of the honeycomb lattice on the layer $n$. 
As illustrated in Fig.~\ref{fig:KHLayer}, the three couplings $J_{\sfx,\sfy,\sfz}$ are assigned for three different bonds on each layer of the honeycomb lattice, and we further assign the $\sfA$- and $\sfB$-links to implement the desired staggered layer structure. 

As is well known, a single layer of the Kitaev honeycomb model produces the $2d$ toric code on the square lattice in the strong-coupling limit of, say, $J_\sfz$ on the $\sfz$-bonds. 
In this limit, we may define effective spin-$1/2$'s on $\sfz$-bonds $\textsf{v} \equiv \langle ij \rangle$ by $\left| + \right>_{\sfv,n} \equiv \left| \uparrow \right>_{i,n} \left| \uparrow \right>_{j,n}$ and $\left| - \right>_{\sfv,n} \equiv \left| \downarrow \right>_{i,n} \left| \downarrow \right>_{j,n}$ where $\sigma^z_{i,n} \left| \uparrow \right>_{i,n} = \left| \uparrow \right>_{i,n}$ and $\sigma^z_{i,n} \left| \downarrow \right>_{i,n} = -\left| \downarrow \right>_{i,n}$. 
These effective spins on each layer now form the square lattice and the spins are defined on the vertices $\sfv$ of the square lattice. 
The $\sfA$- and $\sfB$-links in the Hamiltonian \eqref{eq:StackedKitaevHoneycomb} then belong to each of two sublattices of the square lattice. 
Performing degenerate perturbation theory in the limit $J'_{\sfz \sfz} \to \infty$, we obtain the effective Hamiltonian,
\begin{align} \label{eq:EffHamStackedKH}
H^\textrm{eff}_\textrm{CKH} &= -\frac{J_\sfx^2 J_\sfy^2}{16|J_\sfz|^3} \sum_n \sum_\sfp Y_{\textrm{l}(\sfp),n} Y_{\textrm{r}(\sfp),n} Z_{\textrm{u}(\sfp),n} Z_{\textrm{d}(\sfp),n} \nonumber \\
&-J'_{\sfz \sfz} \sum_m \Biggl( \sum_{\sfv \in \sfA} Z_{\sfv,2m} Z_{\sfv,2m+1} +\sum_{\sfv \in \sfB} Z_{\sfv,2m-1} Z_{\sfv,2m} \Biggr)
\end{align}
where we have defined the Pauli operators $X_{\sfv,n}$, $Y_{\sfv,n}$, and $Z_{\sfv,n}$ acting on the effective spin at the vertex $\sfv$ on the layer $n$. 
The first sum is taken over plaquettes $\sfp$ of the square lattice on each layer, and l($\sfp$), u($\sfp$), r($\sfp$), and d($\sfp$) indicate four corners surrounding the square plaquette $\sfp$ in the clockwise \cite{Kitaev06}. 
This effective Hamiltonian is unitarily equivalent to the model defined in Eq.~\eqref{eq:HamCTC} and one can find that the following unitary transformation does the job:
\begin{align}
(X_{\sfv,n}, Y_{\sfv,n}, Z_{\sfv,n}) \to \begin{cases} (-Y_{\sfv,n}, Z_{\sfv,n}, -X_{\sfv,n}) & \sfv \in \sfA \\ (-Y_{\sfv,n}, X_{\sfv,n}, Z_{\sfv,n}) & \sfv \in \sfB \end{cases}.
\end{align}
We remark that since we have only kept the first-order terms in $J'_{\sfz \sfz}$ in Eq.~\eqref{eq:EffHamStackedKH} the parameters of the original Kitaev honeycomb model must be appropriately tuned such that only the terms remaining in Eq.~\eqref{eq:EffHamStackedKH} are dominant. 
On the other hand, the couplings between adjacent layers are not restrictive to the form in Eq.~\eqref{eq:EffHamStackedKH} but can be, e.g., of the Heisenberg type $\vec{\sigma}_{i,n} \cdot \vec{\sigma}_{i,n+1}$ as far as lowest order perturbations are considered. 

One can also directly work on the strong-coupling limit $J_\sfz = J'_{\sfz \sfz} \to \infty$ and derive the anisotropic fracton model \eqref{eq:AnisotropicFracton} acting on effective qubits $\left| \widetilde{+} \right> = \left| \uparrow \uparrow \uparrow \uparrow \right>$, $\left| \widetilde{-} \right> = \left| \downarrow \downarrow \downarrow \downarrow \right>$ formed by four spins connected by the $J_\sfz$ and $J'_{\sfz \sfz}$ bonds in lowest-order perturbation theory.
Although the Kitaev honeycomb model can be exactly solved by introducing the Majorana representation of spins \cite{Kitaev06}, the additional couplings between layers make the model not solvable and may require some mean-field treatment to fully explore the phase diagram. 

\subsection{$Z_N$ toric code}

\begin{figure}
\includegraphics[clip,width=0.45\textwidth]{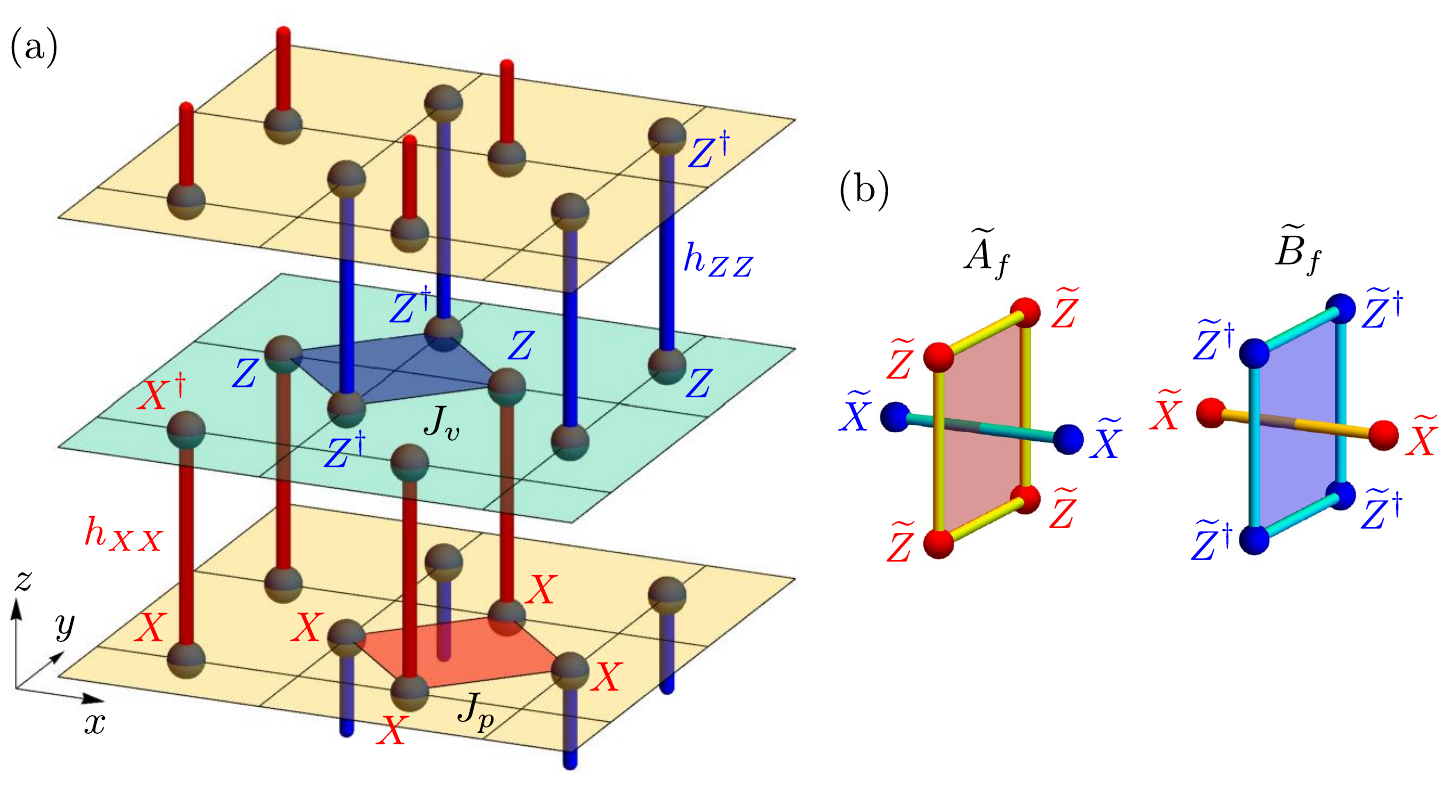}
\caption{(a) Coupled-layer model from the $2d$ $Z_N$ toric code. 
Here, $X$ and $Z$ are the $Z_N$ generalization of the Pauli operators. 
(b) Local terms in the $Z_N$ anisotropic fracton model. 
They are defined on faces of cubes parallel to the $yz$ plane on the bcc lattice as shown in Fig.~\ref{fig:FractonFromTC}~(b).}
\label{fig:ZNLayer}
\end{figure}

It is straightforward to generalize the construction in Sec.~\ref{sec:FractonFromCTC} to coupled layers of the $2d$ $Z_N$ toric code \cite{Kitaev03}. 
Let us define the generalized Pauli operators $X$ and $Z$ satisfying $XZ=\omega ZX$ with $\omega=e^{2\pi i/N}$, which act on the local Hilbert space of a qudit $\left| q \right>$ ($q=0,\cdots,N-1$) as $Z \left| q \right> = \omega^q \left| q \right>$ and $X \left| q \right> = \left| q+1 \mod N \right>$. 
We then consider stacked layers of the $Z_N$ toric code on the square lattice where qudits are defined on each link and introduce couplings between layers to implement anyon condensation. 
In analogy with Eq.~\eqref{eq:HamCTC}, the corresponding Hamiltonian is given by 
\begin{align}
H_{\textrm{C}Z_N} &= \sum_n H_{Z_N,n} \nonumber \\
&-\sum_m \biggl[ h_{XX} \sum_{l \parallel x} (X_{l,2m} X^\dagger_{l,2m+1} +X^\dagger_{l,2m} X_{l,2m+1}) \nonumber \\
&+h_{ZZ} \sum_{l \parallel y} (Z_{l,2m-1} Z^\dagger_{l,2m} +Z^\dagger_{l,2m-1} Z_{l,2m}) \biggr], 
\end{align}
where $H_{Z_N,n}$ is the $Z_N$ toric code Hamiltonian on the $n$-th layer, 
\begin{align}
H_{Z_N,n} = -J_p \sum_p (A_{p,n}+A^\dagger_{p,n}) -J_v \sum_v (B_{v,n}+B^\dagger_{v,n}). 
\end{align}
The local terms $A_{p,n}$ and $B_{v,n}$ are defined by
\begin{align}
\begin{split}
A_{p,n} &= \prod_{l \in p} X_{l,n}, \\
B_{v,n} &=\prod_{\substack{l \in v \\ l \parallel x}} Z_{l,n} \prod_{\substack{l \in v \\ l \parallel y}} Z^\dagger_{l,n}.
\end{split}
\end{align}
The Hamiltonian is schematically shown in Fig.~\ref{fig:ZNLayer}~(a). 

When $h_{XX}=h_{ZZ}=0$, the Hamiltonian is just decoupled layers of the $Z_N$ toric code, which is exactly solvable. 
As the Hamiltonian consists of local commuting terms and $A_{p,n}^N = B_{v,n}^N=1$, the ground state is given by a simultaneous eigenstate of $A_{p,n}$ and $B_{v,n}$ with the eigenvalues $+1$. 
By acting an $X$ operator on the ground state, it creates a pair of $e$ and $e^{N-1}$ excitations on neighboring vertices, and similarly $Z$ creates a pair of $m$ and $m^{N-1}$ on neighboring plaquettes. 
Each layer hosts quasiparticles of the $Z_N$ topological order given by $\{ \id, e, \cdots, e^{N-1} \}$ and $\{ \id, m, \cdots, m^{N-1} \}$ and their bound objects. 
Here, $e$ and $m$ are bosons but have the nontrivial mutual statistics of $2\pi/N$; 
the quasiparticle $e^q$ or $m^q$ ($q=2,\cdots,N-1$) are obtained by fusing $q$ $e$'s or $m$'s, respectively, while fusing $N$ $e$'s or $m$'s leads to the vacuum $\id$. 

A finite coupling $h_{XX}$ will induce the condensation of $e$-$e^{N-1}$ pairs between adjacent layers, while $h_{ZZ}$ will induce the condensation of $m$-$m^{N-1}$ pairs. 
A single $e$ excitation created on a vertex of a layer can propagate through the $e$-$e^{N-1}$ condensate along the $x$ axis, and similarly a single $m$ created on a plaquette propagate through the $m$-$m^{N-1}$ condensate along the $x$ axis; they behave as lineon excitations. 
On the other hand, the $e$ excitation sees the $m$-$m^{N-1}$ condensate in the $y$ direction, and it must form an $e$-$e$ pair with the neighboring layer to move along the $y$ axis. 
Similarly, only an $m$-$m$ pair can move along the $y$ axis. 
Thus, dipoles of $e$'s or $m$'s separated along the $z$ axis behave as planon excitations in the $xy$ plane. 
Finally, when $e$-$e$ pairs and $e^{N-1}$-$e^{N-1}$ pairs are alternatively created over several layers, the internal $e$-$e^{N-1}$ pairs become invisible in the $e$-$e^{N-1}$ condensate and only dipoles of $e$ and $e^{N-1}$ separated along the $y$ axis are left at the top and bottom layers. 
This dipole nature similarly applies to $m$ excitations. 
Such dipoles behave as planons in the $xz$ plane. 

This phenomenology from anyon condensation can be explicitly seen in the effective Hamiltonian obtained in the strong-coupling limit $h_{XX}, h_{ZZ} \to \infty$. 
In this limit, two qudits on a link parallel to the $x$ axis are in one of the following $N$ states labeled by $q=0,\cdots,N-1$, 
\begin{align}
\left| \tq \right>_{l,2m+1/2} \equiv \left| q \right>_{l,2m} \left| q \right>_{l,2m+1}, 
\end{align}
in the basis satisfying $X_{l,n} \left| q \right>_{l,n} = \omega^q \left| q \right>_{l,n}$, while for those on a link parallel to the $y$ axis we have
\begin{align}
\left| \tq \right>_{l,2m-1/2} \equiv \left| q \right>_{l,2m-1} \left| q \right>_{l,2m}
\end{align}
in the basis satisfying $Z_{l,n} \left| q \right>_{l,n} = \omega^q \left| q \right>_{l,n}$. 
As in Sec.~\ref{sec:StrongCouplingAF}, the effective qudits $\left| \tq \right>$ live on the bcc lattice. 
We can then perform degenerate perturbation theory and obtain an effective Hamiltonian acting on the Hilbert space of the effective qudits, 
\begin{align} \label{eq:ZNAnisotropicFracton}
H^\textrm{eff}_{\textrm{C}Z_N} &= -\frac{J_p^2}{8h_{ZZ} \sin^2 (\pi/N)} \sum_{\substack{f \in A \\ f \parallel yz}} (\tA_f +\tA_f^\dagger) \nonumber \\
&-\frac{J_v^2}{8h_{XX} \sin^2 (\pi/N)} \sum_{\substack{f \in B \\ f \parallel yz}} (\tB_f +\tB_f^\dagger),
\end{align}
where the local terms $\tA_f$ and $\tB_f$ are defined by
\begin{align}
\begin{split}
\tA_f &= \prod_{s \in f} \tZ^A_s \prod_{s' \perp f} \tX^B_{s'}, \\
\tB_f &= \prod_{s \in f} \tZ^{B\dagger}_s \prod_{s' \perp f} \tX^A_{s'},
\end{split}
\end{align}
as shown in Fig.~\ref{fig:ZNLayer}~(b).
Here, we have defined generalized Pauli operators, $\tX^{A(B)}_s$ and $\tZ^{A(B)}_s$, acting on an effective qudit at the site $s$ on the $A$ ($B$) sublattice of the bcc lattice: $\tZ \left| \tq \right> = \omega^q \left| \tq \right>$ and $\tX \left| \tq \right> = \left| \tq +1 \mod N \right>$. 
The Hamiltonian \eqref{eq:ZNAnisotropicFracton} has also been introduced in Ref.~\cite{Shirley18a} and is again exactly solvable. 
The ground state admits lineon and planon excitations as discussed above from the view of anyon condensation, resulting in the ground-state degeneracy $N^{2(L_y+L_z-1)}$ on a torus.

\subsection{Toric code on the honeycomb lattice}
\label{sec:TCHoneycomb}

\begin{figure}
\includegraphics[clip,width=0.45\textwidth]{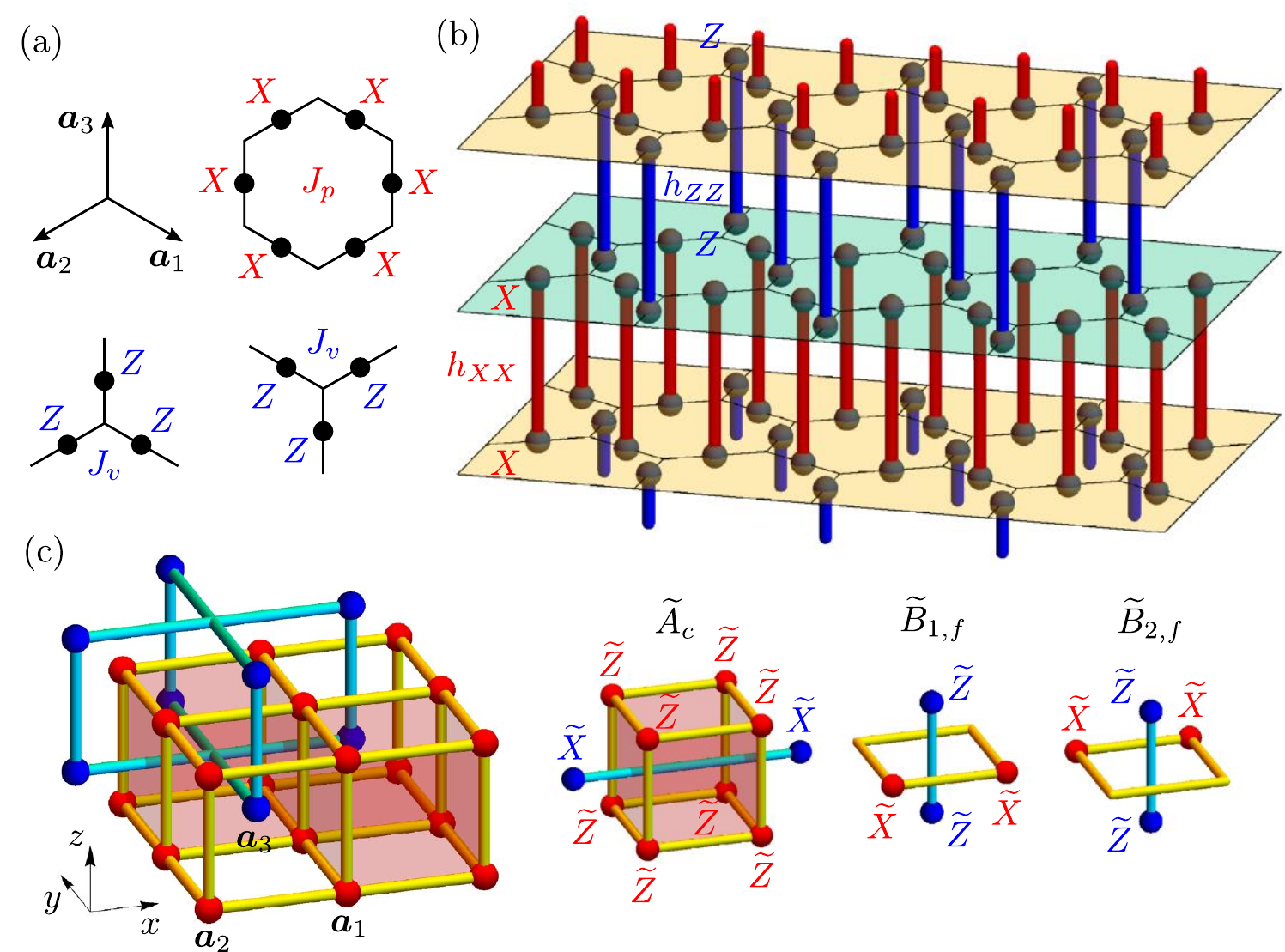}
\caption{(a) Local terms in the toric code Hamiltonian on the honeycomb lattice. 
We also define the vectors $\bfa_{1,2,3}$ to specify three different links in the $xy$ plane. 
(b) Coupled-layer model from the $2d$ toric codes on the honeycomb lattice. 
The red (blue) bonds represent the $h_{XX}$ ($h_{ZZ}$) couplings between $\bfa_{1,2}$ ($\bfa_3$) links on adjacent layers. 
(c) Effective Hamiltonian obtained in the strong-coupling limit. 
There are three types of local terms $\tA_c$, $\tB_{1,f}$, and $\tB_{2,f}$ acting on effective qubits.}
\label{fig:TCHoneycombLayer}
\end{figure}

We can generalize the coupled-layer construction of the anisotropic fracton model in Sec.~\ref{sec:FractonFromCTC} to the $2d$ toric codes defined on trivalent graphs. 
Here, let us specifically focus on the honeycomb lattice. 
The coupled-layer model is defined by the Hamiltonian, 
\begin{align} \label{eq:HamCTCh}
H_{\textrm{CTCh}} &= \sum_n H_{\textrm{TCh},n} -\sum_m \biggl( h_{XX} \sum_{l \parallel \bfa_1} X_{l,2m} X_{l,2m+1} \nonumber \\
&+h_{XX} \sum_{l \parallel \bfa_2} X_{l,2m} X_{l,2m+1} \nonumber \\
&+h_{ZZ} \sum_{l \parallel \bfa_3} Z_{l,2m-1} Z_{l,2m} \biggr),
\end{align}
where $H_{\textrm{TCh},n}$ is the toric code Hamiltonian on the $n$-th layer,
\begin{align} \label{eq:HamTCh}
H_{\textrm{TCh},n} = -J_p \sum_p \prod_{l \in p} X_{l,n} -J_v \sum_v \prod_{l \in v} Z_{l,n}.
\end{align}
Here, the Pauli operators $X_{l,n}$ and $Z_{l,n}$ act on a qubit placed on the link $l$ of the honeycomb lattice on the $n$-th layer. 
Since there are three inequivalent links under $120^\circ$ rotation, we specify them by the vectors $\bfa_{1,2,3}$ as defined in Fig.~\ref{fig:TCHoneycombLayer}~(a). 
The local terms in Eq.~\eqref{eq:HamTCh} act on plaquettes with six links or vertices with three links as also shown in Fig.~\ref{fig:TCHoneycombLayer}~(a).
Similarly to the case for the square lattice, we implement a staggered structure of coupling between layers; links parallel to $\bfa_1$ or $\bfa_2$ are coupled via $XX$ between the $2m$-th and $(2m+1)$-th layers, while those parallel to $\bfa_3$ are coupled via $ZZ$ between the $(2m-1)$-th and $2m$-th layers, as shown in Fig.~\ref{fig:TCHoneycombLayer}~(b). 
We then perform degenerate perturbation theory in the strong-coupling limit $h_{XX}, h_{ZZ} \to \infty$ and obtain the effective Hamiltonian for effective qubits on the vertical bonds,
\begin{align} \label{eq:AnisotropicFractonHoneycomb}
H^\textrm{eff}_{\textrm{CTCh}} = -\frac{J_p^2}{4h_{ZZ}} \sum_c \tA_c -\frac{J_v^2}{4h_{XX}} \sum_f \left( \tB_{1,f} +\tB_{2,f} \right).
\end{align}
We here regard that the effective qubits on links parallel to $\bfa_1$ or $\bfa_2$ form a cubic lattice as shown in Fig.~\ref{fig:TCHoneycombLayer}~(c). 
The effective qubits on links parallel to $\bfa_3$ live on the center of cubes to form a checkerboard pattern in the $xy$ plane.
The first sum in Eq.~\eqref{eq:AnisotropicFractonHoneycomb} is taken over cubes $c$ that do not contain the qubits on the $\bfa_3$ links, whereas the second sum is taken over faces $f$ that are parallel to the $xy$ plane and  belong to cubes containing the qubits on the $\bfa_3$ links. 
The local operators $\tA_c$, $\tB_{1,f}$, and $\tB_{2,f}$ are written in terms of products of Pauli operators $\tX$ and $\tZ$ acting on the effective qubits, whose specific forms are pictorially shown in Fig.~\ref{fig:TCHoneycombLayer}~(c). 

\begin{figure}
\includegraphics[clip,width=0.45\textwidth]{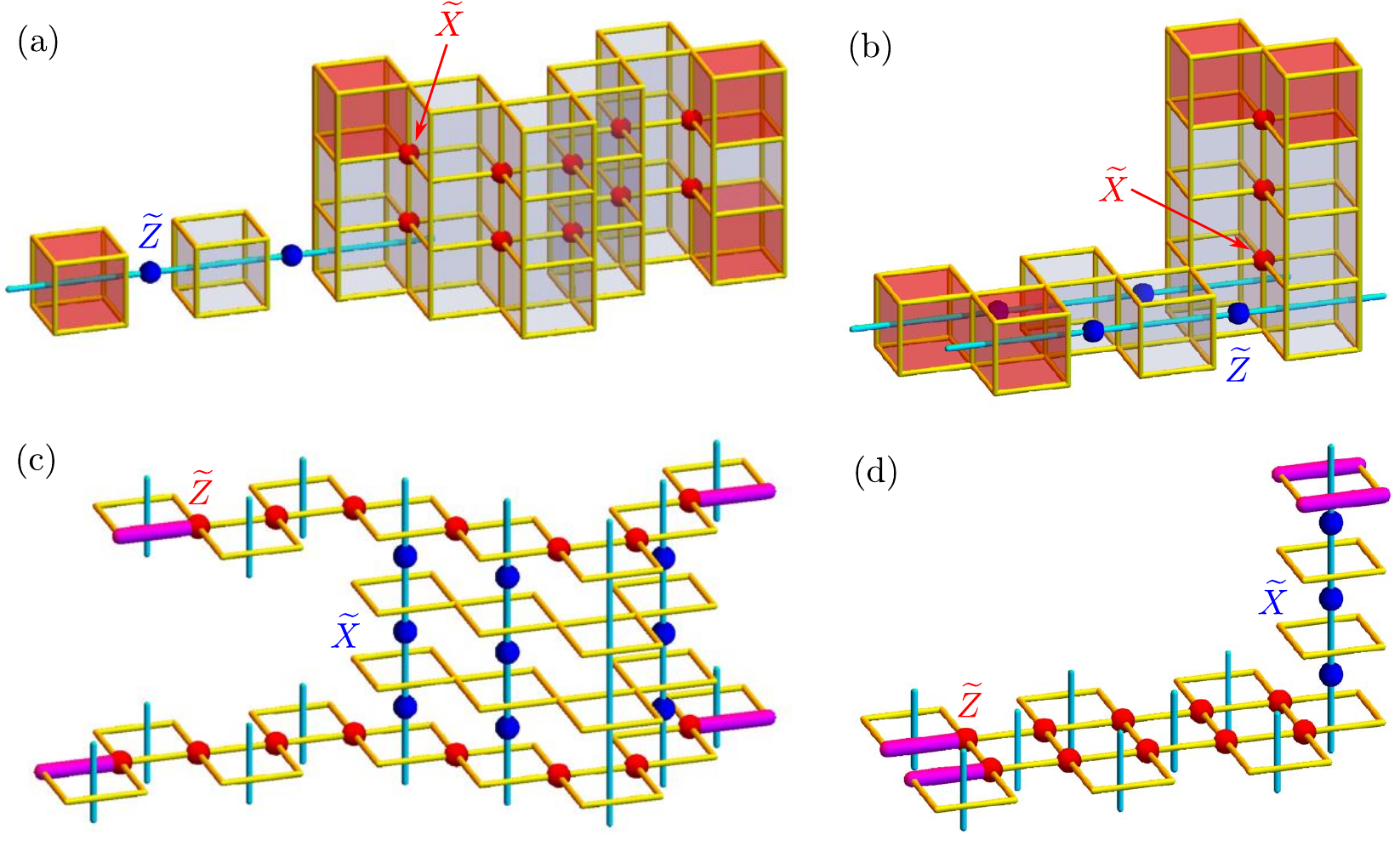}
\caption{Subdimensional excitations in the coupled toric-code model on the honeycomb lattice. 
(a)-(b) Lineon excitations created by flipping the eigenvalues of $\tA_c$ are marked by the red cubes.
Their dipole separated along the $z$ axis behave as a planon in the $xy$ plane (a), while those separated within the $xy$ plane behave as a planon in the $xz$ plane (b). 
(c)-(d) Lineon excitations created by flipping the eigenvalues of $\tB_{1,f}$ and $\tB_{2,f}$ are marked by the magenta bonds. 
Their dipole similarly behave as a planon in the $xy$ plane (c) or in the $xz$ plane (d).}
\label{fig:ExcitationsFTCH}
\end{figure}

The Hamitonian \eqref{eq:AnisotropicFractonHoneycomb} is a commuting projector Hamiltonian and thus is exactly solvable. 
As we demonstrate below, it is a variant of the anisotropic fracton model considered above.
We define the unit cell to contain three qubits from the $\bfa_{1,2,3}$ links and consider the $L_x \times L_y \times L_z$ torus with $3L_x L_y L_z$ effective qubits. 
As similarly seen in Sec.~\ref{sec:GSD}, the $3L_x L_y L_z$ local operators in the Hamiltonian do not fully span the Hilbert space and turn out to leave the ground-state degeneracy, 
\begin{align}
\textrm{GSD} = 2^{2(L_y+L_z-1)}, 
\end{align}
since the operators $\tA_c$ are multiplied over the $xy$ or $xz$ plane to be identity and similarly for $\tB_{1,f}$ and $\tB_{2,f}$. 
This subextensive degeneracy is again a consequence of subdimensional excitations. 
Successively acting Pauli $\tZ$ operators on $\bfa_3$ links on a straight line along the $x$ axis, they flip the eigenvalues of $\tA_c$ on the line and create lineon excitations. 
However, their dipoles separated along the $z$ axis can move on the $xy$ plane by acting Pauli $\tX$ operators on $\bfa_{1,2}$ links and thus become planon excitations, as illustrated in Fig.~\ref{fig:ExcitationsFTCH}~(a). 
On the other hand, dipoles created within the $xy$ plane become planon excitations moving on the $xz$ plane [Fig.~\ref{fig:ExcitationsFTCH}~(b)]. 
Acting Pauli $\tX$ operators on $\bfa_{1,2}$ links along the $x$ axis, they flip the eigenvalues of $\tB_{1,f}$ and $\tB_{2,f}$ and create another type of lineon excitations. 
With the action of Pauli $\tX$ operators on $\bfa_3$ links, their dipoles behave as planon excitations in the $xy$ plane when they are separated along the $z$ axis [Fig.~\ref{fig:ExcitationsFTCH}~(c)] or in the $xz$ plane when they are separated along the $y$ axis [Fig.~\ref{fig:ExcitationsFTCH}~(d)]. 
These anisotropic behaviors of quasiparticle excitations can also be understood from the view of anyon condensation as discussed in Sec.~\ref{sec:AFAnyonCondens} by taking account of the lattice and coupling structure accordingly.

It appears that we can obtain another anisotropic fracton model by interchanging the $XX$ and $ZZ$ couplings in Eq.~\eqref{eq:HamCTCh} and by taking the strong-coupling limit, although this makes just a $90^\circ$ rotation for the coupled-layer model defined on the square lattice in Sec.~\ref{sec:FractonFromCTC}. 
However, the degenerate perturbation theory generates local terms consisting of plaquette and vertex operators not only between adjacent layers but also within the same layer at the second order. 
These terms render the resulting model trivial up to a stack of $2d$ topological orders. 
As discussed in Sec.~\ref{sec:Conclusion}, this might be a generic feature of our construction when the coupling between layers makes the original quasiparicles (single $e$'s or $m$'s in the toric code) immobile within a layer.

\subsection{Doubled semion model}

\begin{figure}
\includegraphics[clip,width=0.45\textwidth]{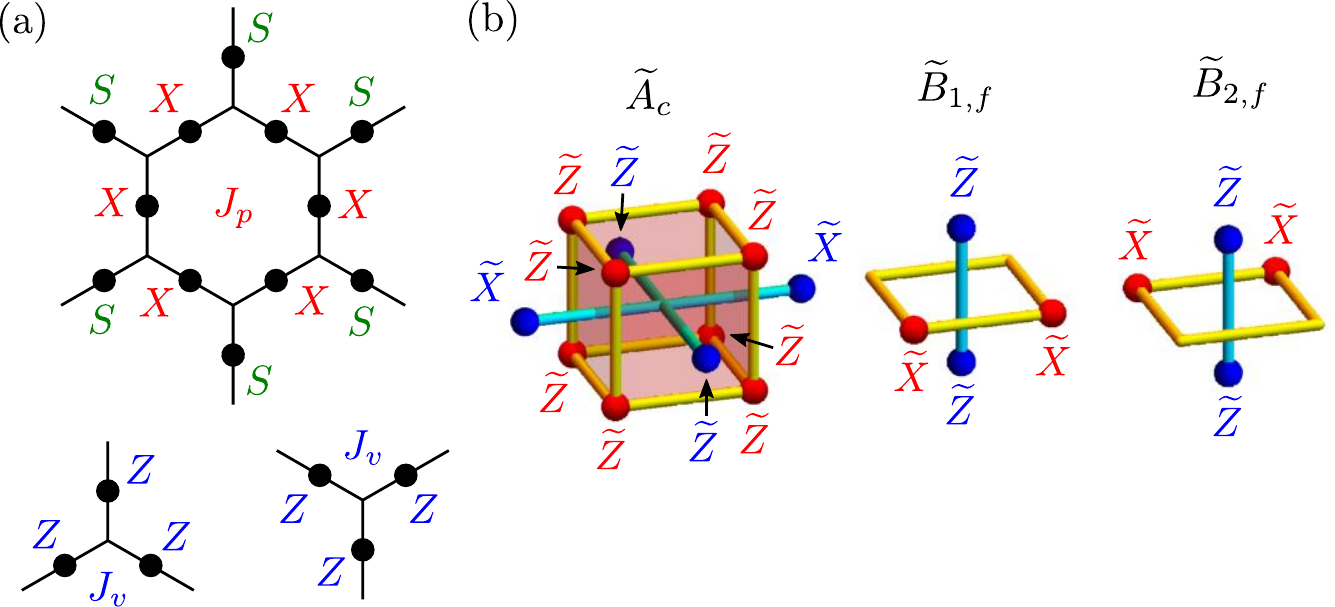}
\caption{(a) Local terms in the doubled semion model. 
The plaquette term is modified from the toric code in such a way that it involves the operators $S=\textrm{diag}(1,i)$ on six links coming into the plaquette and the projection operators $P_v$ on six verices. 
(b) Local terms in the effective Hamiltonian obtained from the coupled-layer model. 
The cube term $\tA_c$ is now modified from that in Fig.~\ref{fig:TCHoneycombLayer}~(c) to involve the Pauli $\tZ$ operators at the ends of a bond parallel to the $y$ axis.}
\label{fig:DoubledSemion}
\end{figure}

As we have constructed an anisotropic fracton model from coupled layers of the toric code on the honeycomb lattice, there might be generalizations to the string-net models \cite{Levin05}, which provide exactly solvable Hamiltonians for various $2d$ topological orders on trivalent graphs including the honeycomb lattice. 
One of the simplest models among them is the doubled semion model, which has been used for the coupled-layer construction of the semionic X-cube model \cite{HMa17}. 
We thus consider coupled layers of the doubled semion model on the honeycomb lattice, 
\begin{align} \label{eq:HamCDS}
H_{\textrm{CDS}} &= \sum_n H_{\textrm{DS},n} -\sum_m \biggl[ h_{XX} \sum_{l \parallel \bfa_1} X_{l,2m} X_{l,2m+1} \nonumber \\
&+h_{XX} \sum_{l \parallel \bfa_2} X_{l,2m} X_{l,2m+1} +h_{ZZ} \sum_{l \parallel \bfa_3} Z_{l,2m-1} Z_{l,2m} \biggr]. 
\end{align}
Here, $H_{\textrm{DS},n}$ is the Hamiltonian for the doubled semion model on the $n$-th layer,
\begin{align}
H_{\textrm{DS},n} &= -J_p \sum_p \prod_{l \in p} X_{l,n} \prod_{l \perp p} S_{l,n} \prod_{v \in p} P_{v,n} \nonumber \\
&-J_v \sum_v \prod_{l \in v} Z_{l,n}, 
\end{align}
where the first term consists of the Pauli $X$ operators acting on six links of the plaquette $p$, the operators $S=\textrm{diag}(1,i)$ acting on six links coming into the plaquette, and the projection operators $P_{v,n}$ acting on six vertices $v$ of the plaquette to enforce the constraints on each vertex $\prod_{l \in v} Z_{l,n}=1$ [see Fig.~\ref{fig:DoubledSemion}~(a)]. 
We have implemented the same structure of coupling between layers as in the case for the toric code. 

Similarly to the toric code, a Pauli $Z$ operator flips the eigenvalues of plaquette terms and thus create excitations on the plaquettes. 
These excitations can be seen as bosonic quasiparticles ($b$) in the doubled semion topological order \cite{Levin05}. 
Therefore, the $ZZ$ coupling in Eq.~\eqref{eq:HamCDS} will induce the condensation of $bb$ between adjacent layers. 
On the other hand, a Pauli $X$ operator does not solely flip the eigenvalues of vertex terms for the doubled semion model. 
In fact, the corresponding string operator must be appropriately modified to selectively excite semion ($s$) or anti-semion ($\bar{s}$) excitations \cite{Levin05}, and the action of a single $X$ operator rather yields a superposition of excited states with $s$ or $\bar{s}$. 
Thus, we cannot draw a clear picture from the anyon condensation induced by the $XX$ coupling to discuss the motion of quasiparticles in this case. 
Nevertheless, we can write down the effective Hamiltonian in the strong-coupling limit $h_{XX},h_{ZZ} \to \infty$, which takes the same form as Eq.~\eqref{eq:AnisotropicFractonHoneycomb} at the second-order with modified cube terms $\tA_c$ as illustrated in Fig.~\ref{fig:DoubledSemion}~(b). 
It still keeps a commuting-projector form and is exactly solvable. 
The ground-state degeneracy on the torus is also not changed from $2^{2(L_y+L_z-1)}$. 
Although the action of a single $\tX$ operator is changed as it creates excitations in $\tA_c$ as well as $\tB_{1,f}$ and $\tB_{2,f}$, causing a slightly complicated behavior for dipole excitations, the model possesses quasiparticle excitations in the forms of lineons and planons essentially similar to those discussed in Sec.~\ref{sec:TCHoneycomb}. 
Hence, what we have obtained from the doubled semion model is still a variant of the anisotropic fracton model.

\section{Conclusion and discussion}
\label{sec:Conclusion}

In this work, we have proposed a coupled-layer construction of anisotropic fracton models from layers of $2d$ topological orders stacked in one spatial direction. 
Quasiparticle excitations in fracton phases have been studied by analyzing the effective Hamiltonians in the strong-coupling limit or by considering the pair condensation of quasiparticles of the original $2d$ topological orders between adjacent layers. 
The subdimensional excitations show the spatially anisotropic mobility of lineons moving only along a straight line or planons moving only within a $2d$ plane, depending on how the original quasiparticles on each layer see the anyon condensates induced by the coupling between layers. 

\begin{figure}
\includegraphics[clip,width=0.48\textwidth]{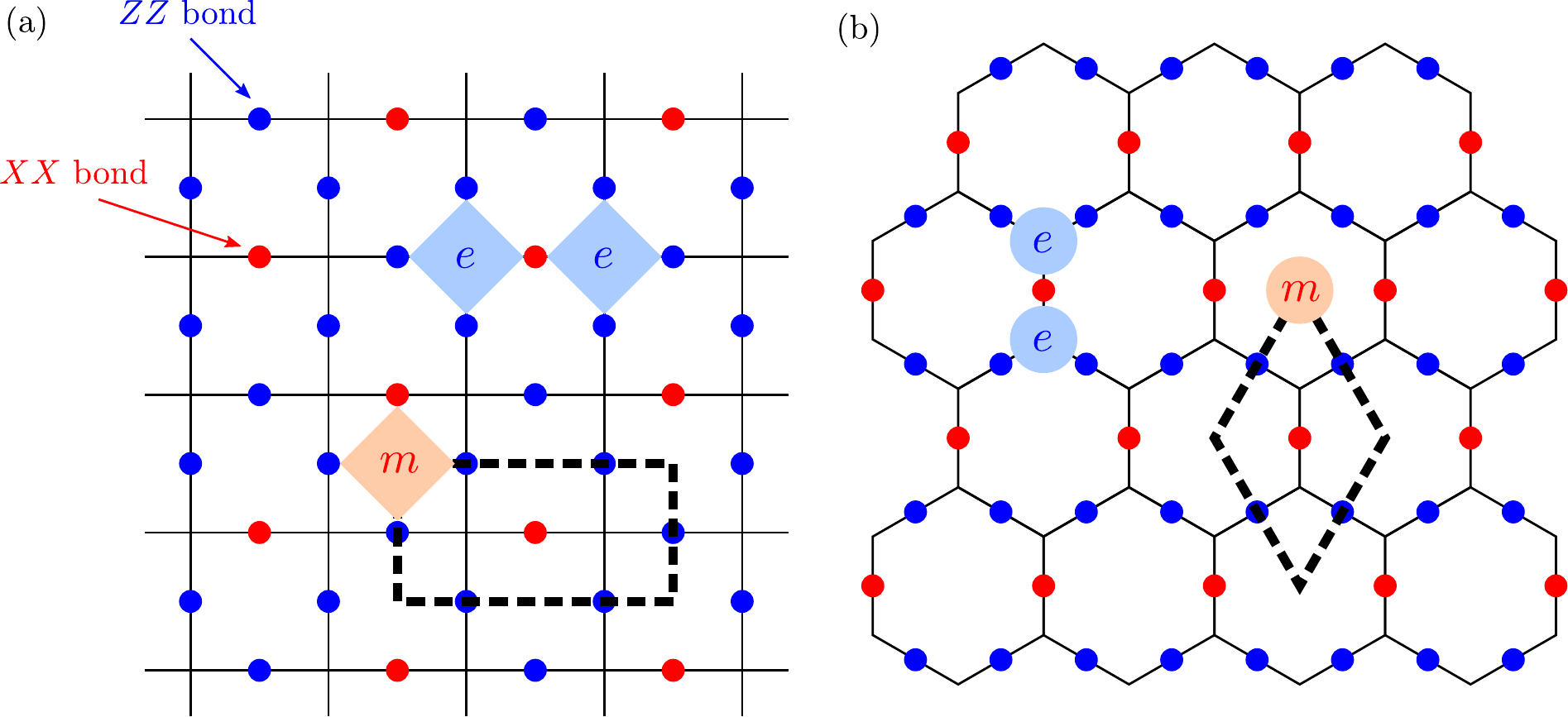}
\caption{Patterns of anyon condensation that gives trivial models on (a) the square lattice and (b) the honeycomb lattice. 
Similarly to Fig.~\ref{fig:AnyonCondensAF}~(a), we take a slice of the coupled-layer models.
In both cases, single $e$ excitations are trapped at the $XX$ bonds, while single $m$ excitations can move along a closed loop (dashed line) around the $XX$ bond.}
\label{fig:TrivialAnyonCondens}
\end{figure}

We can then ask what happens when the anyon condensation between layers do not allow the original quasiparticles to move in isolation---the situation reminiscent of immobile, fracton excitations. 
For instance, let us consider two patterns of the $XX$ and $ZZ$ coupling between layers of the $2d$ toric codes defined on the square or honeycomb lattice, as depicted in Fig.~\ref{fig:TrivialAnyonCondens}.
In such cases, either single $e$ or $m$ excitations cannot move, while another kind of excitations can move within a $2d$ layer and can form closed loops around where the other excitations are trapped. 
Performing degenerate perturbation theory in the strong-coupling limit, we have two types of nonvanishing terms: 
One consists of the original plaquette and vertex terms between adjacent layers, while another consists of those within the same layer; the latter results from the presence of excitations moving along the loops. 
If we discard such intralayer terms in the effective Hamiltonian, the ground state on a torus has degeneracy exponentially scaling with $L_x L_y$, which is the number of closed loops. 
However, this degeneracy is lifted once the intralayer terms are added, leaving only the degeneracy coming from $2d$ topological orders stacked along the $z$ direction. 
Thus, the resulting models become trivial as fracton topological order. 
In order to have nontrivial fracton models, the anyon condensation will necessarily be implemented in such a way that both $e$ and $m$ excitations can move along some one-dimensional paths. 
Such models will have lineons and planons as dipoles of the lineons, but no fractons. 
We expect that this is a generic feature of our construction using $2d$ topological orders stacked only along one direction.

There are several other questions naturally raised: 
(i) Although we did not address the nature of transitions from decoupled $2d$ topological orders to anisotropic fracton models, the duality mapping \cite{Vijay17a, Slagle17a} suggests that the transition for the $Z_2$ case is described by a stack of transverse Ising chains. 
Once those chains are coupled, the transition is expected to be first order or accompanied with an intermediate phase \cite{Slagle17a}. 
Since the models are simpler than those obtained from the previous layer constructions \cite{HMa17, Vijay17a, Vijay17b, Slagle17a, Prem19a, Shirley19d}, direct numerical simulations may be more capable of determining the nature of the phase transitions. 
(ii) As a model has been obtained from a single stack of layers of the Kitaev honeycomb models in Sec.~\ref{sec:KitaevHoneycomb}, it is interesting to seek possibilities to realize fracton orders in spin-orbit-coupled materials in the proximity to a $2d$ spin liquid but with non-negligible $3d$ couplings \cite{Jackeli09, Chaloupka10}. 
(iii) There may be a generalization of the present layer construction for twisted cousins of the anisotropic fracton models as proposed for the X-cube model and others \cite{YYou18a, HSong19, Shirley19d}. 

The construction using $2d$ topologically ordered phases stacked only in one direction bears a resemblance to the layer construction proposed for conventional $3d$ topological orders or surface topological orders of $3d$ topological phases \cite{CMJian14}. 
The simplest example of the latter construction requires the successive condensation of $ee$ pairs between layers of the $2d$ toric codes and is expected to give the $3d$ $Z_2$ topological order with bosonic point-like quasiparticles of $e$ and loop-like quasiparticles made of strings of $m$'s. 
Such pair condensations could be naively implemented in a lattice model by coupling many layers of the $2d$ toric codes via $XX$ terms between neighboring layers as similarly done in Sec.~\ref{sec:CondensTwoLayers}. 
However, deconfined quasiparticles made of $m$'s appear to take only the form of straight strings threading all layers but not closed loops, as already raised as a subtlety in the original work. 
On the other hand, the author, in a collaboration with Furusaki, has recently achieved a successful implementation of this phenomenology in coupled layers further made of coupled quantum wires \cite{Fuji19} (see also Ref.~\cite{Iadecola19}). 
Analogous coupled-wire constructions may be possible for fracton topological orders and might be generalized to stacked layers of $2d$ topological orders that do not admit the lattice realization in commuting projector Hamiltonians, such as chiral topological orders. 
It is left for a future work.

\section*{Acknowledgment}
The author thanks Yasuyuki Kato and Tomohiro Soejima for valuable discussions, and Kevin Slagle for useful comments on the draft, especially about the duality mapping to investigate phase transitions.

\bibliography{AnisotropicFracton}

\begin{thebibliography}{85}%
\makeatletter
\providecommand \@ifxundefined [1]{%
 \@ifx{#1\undefined}
}%
\providecommand \@ifnum [1]{%
 \ifnum #1\expandafter \@firstoftwo
 \else \expandafter \@secondoftwo
 \fi
}%
\providecommand \@ifx [1]{%
 \ifx #1\expandafter \@firstoftwo
 \else \expandafter \@secondoftwo
 \fi
}%
\providecommand \natexlab [1]{#1}%
\providecommand \enquote  [1]{``#1''}%
\providecommand \bibnamefont  [1]{#1}%
\providecommand \bibfnamefont [1]{#1}%
\providecommand \citenamefont [1]{#1}%
\providecommand \href@noop [0]{\@secondoftwo}%
\providecommand \href [0]{\begingroup \@sanitize@url \@href}%
\providecommand \@href[1]{\@@startlink{#1}\@@href}%
\providecommand \@@href[1]{\endgroup#1\@@endlink}%
\providecommand \@sanitize@url [0]{\catcode `\\12\catcode `\$12\catcode
  `\&12\catcode `\#12\catcode `\^12\catcode `\_12\catcode `\%12\relax}%
\providecommand \@@startlink[1]{}%
\providecommand \@@endlink[0]{}%
\providecommand \url  [0]{\begingroup\@sanitize@url \@url }%
\providecommand \@url [1]{\endgroup\@href {#1}{\urlprefix }}%
\providecommand \urlprefix  [0]{URL }%
\providecommand \Eprint [0]{\href }%
\providecommand \doibase [0]{http://dx.doi.org/}%
\providecommand \selectlanguage [0]{\@gobble}%
\providecommand \bibinfo  [0]{\@secondoftwo}%
\providecommand \bibfield  [0]{\@secondoftwo}%
\providecommand \translation [1]{[#1]}%
\providecommand \BibitemOpen [0]{}%
\providecommand \bibitemStop [0]{}%
\providecommand \bibitemNoStop [0]{.\EOS\space}%
\providecommand \EOS [0]{\spacefactor3000\relax}%
\providecommand \BibitemShut  [1]{\csname bibitem#1\endcsname}%
\let\auto@bib@innerbib\@empty
\bibitem [{\citenamefont {Wen}(2016)}]{Wen16}%
  \BibitemOpen
  \bibfield  {author} {\bibinfo {author} {\bibfnamefont {X.-G.}\ \bibnamefont
  {Wen}},\ }\href {\doibase 10.1093/nsr/nwv077} {\bibfield  {journal} {\bibinfo
   {journal} {Nat. Sci. Rev.}\ }\textbf {\bibinfo {volume} {3}},\ \bibinfo
  {pages} {68} (\bibinfo {year} {2016})}\BibitemShut {NoStop}%
\bibitem [{\citenamefont {Hamma}\ \emph {et~al.}(2005)\citenamefont {Hamma},
  \citenamefont {Zanardi},\ and\ \citenamefont {Wen}}]{Hamma05}%
  \BibitemOpen
  \bibfield  {author} {\bibinfo {author} {\bibfnamefont {A.}~\bibnamefont
  {Hamma}}, \bibinfo {author} {\bibfnamefont {P.}~\bibnamefont {Zanardi}}, \
  and\ \bibinfo {author} {\bibfnamefont {X.-G.}\ \bibnamefont {Wen}},\ }\href
  {\doibase 10.1103/PhysRevB.72.035307} {\bibfield  {journal} {\bibinfo
  {journal} {Phys. Rev. B}\ }\textbf {\bibinfo {volume} {72}},\ \bibinfo
  {pages} {035307} (\bibinfo {year} {2005})}\BibitemShut {NoStop}%
\bibitem [{\citenamefont {Wang}\ and\ \citenamefont
  {Levin}(2014)}]{ChenjieWang14}%
  \BibitemOpen
  \bibfield  {author} {\bibinfo {author} {\bibfnamefont {C.}~\bibnamefont
  {Wang}}\ and\ \bibinfo {author} {\bibfnamefont {M.}~\bibnamefont {Levin}},\
  }\href {\doibase 10.1103/PhysRevLett.113.080403} {\bibfield  {journal}
  {\bibinfo  {journal} {Phys. Rev. Lett.}\ }\textbf {\bibinfo {volume} {113}},\
  \bibinfo {pages} {080403} (\bibinfo {year} {2014})}\BibitemShut {NoStop}%
\bibitem [{\citenamefont {Jiang}\ \emph {et~al.}(2014)\citenamefont {Jiang},
  \citenamefont {Mesaros},\ and\ \citenamefont {Ran}}]{SJiang14}%
  \BibitemOpen
  \bibfield  {author} {\bibinfo {author} {\bibfnamefont {S.}~\bibnamefont
  {Jiang}}, \bibinfo {author} {\bibfnamefont {A.}~\bibnamefont {Mesaros}}, \
  and\ \bibinfo {author} {\bibfnamefont {Y.}~\bibnamefont {Ran}},\ }\href
  {\doibase 10.1103/PhysRevX.4.031048} {\bibfield  {journal} {\bibinfo
  {journal} {Phys. Rev. X}\ }\textbf {\bibinfo {volume} {4}},\ \bibinfo {pages}
  {031048} (\bibinfo {year} {2014})}\BibitemShut {NoStop}%
\bibitem [{\citenamefont {Lin}\ and\ \citenamefont {Levin}(2015)}]{CHLin15}%
  \BibitemOpen
  \bibfield  {author} {\bibinfo {author} {\bibfnamefont {C.-H.}\ \bibnamefont
  {Lin}}\ and\ \bibinfo {author} {\bibfnamefont {M.}~\bibnamefont {Levin}},\
  }\href {\doibase 10.1103/PhysRevB.92.035115} {\bibfield  {journal} {\bibinfo
  {journal} {Phys. Rev. B}\ }\textbf {\bibinfo {volume} {92}},\ \bibinfo
  {pages} {035115} (\bibinfo {year} {2015})}\BibitemShut {NoStop}%
\bibitem [{\citenamefont {Wang}\ and\ \citenamefont {Wen}(2015)}]{JCWang15}%
  \BibitemOpen
  \bibfield  {author} {\bibinfo {author} {\bibfnamefont {J.~C.}\ \bibnamefont
  {Wang}}\ and\ \bibinfo {author} {\bibfnamefont {X.-G.}\ \bibnamefont {Wen}},\
  }\href {\doibase 10.1103/PhysRevB.91.035134} {\bibfield  {journal} {\bibinfo
  {journal} {Phys. Rev. B}\ }\textbf {\bibinfo {volume} {91}},\ \bibinfo
  {pages} {035134} (\bibinfo {year} {2015})}\BibitemShut {NoStop}%
\bibitem [{\citenamefont {Vijay}\ \emph {et~al.}(2016)\citenamefont {Vijay},
  \citenamefont {Haah},\ and\ \citenamefont {Fu}}]{Vijay16}%
  \BibitemOpen
  \bibfield  {author} {\bibinfo {author} {\bibfnamefont {S.}~\bibnamefont
  {Vijay}}, \bibinfo {author} {\bibfnamefont {J.}~\bibnamefont {Haah}}, \ and\
  \bibinfo {author} {\bibfnamefont {L.}~\bibnamefont {Fu}},\ }\href {\doibase
  10.1103/PhysRevB.94.235157} {\bibfield  {journal} {\bibinfo  {journal} {Phys.
  Rev. B}\ }\textbf {\bibinfo {volume} {94}},\ \bibinfo {pages} {235157}
  (\bibinfo {year} {2016})}\BibitemShut {NoStop}%
\bibitem [{\citenamefont {Chamon}(2005)}]{Chamon05}%
  \BibitemOpen
  \bibfield  {author} {\bibinfo {author} {\bibfnamefont {C.}~\bibnamefont
  {Chamon}},\ }\href {\doibase 10.1103/PhysRevLett.94.040402} {\bibfield
  {journal} {\bibinfo  {journal} {Phys. Rev. Lett.}\ }\textbf {\bibinfo
  {volume} {94}},\ \bibinfo {pages} {040402} (\bibinfo {year}
  {2005})}\BibitemShut {NoStop}%
\bibitem [{\citenamefont {Bravyi}\ \emph {et~al.}(2011)\citenamefont {Bravyi},
  \citenamefont {Leemhuis},\ and\ \citenamefont {Terhal}}]{Bravyi11}%
  \BibitemOpen
  \bibfield  {author} {\bibinfo {author} {\bibfnamefont {S.}~\bibnamefont
  {Bravyi}}, \bibinfo {author} {\bibfnamefont {B.}~\bibnamefont {Leemhuis}}, \
  and\ \bibinfo {author} {\bibfnamefont {B.~M.}\ \bibnamefont {Terhal}},\
  }\href {\doibase 10.1016/j.aop.2010.11.002} {\bibfield  {journal} {\bibinfo
  {journal} {Ann. Phys.}\ }\textbf {\bibinfo {volume} {326}},\ \bibinfo {pages}
  {839 } (\bibinfo {year} {2011})}\BibitemShut {NoStop}%
\bibitem [{\citenamefont {Haah}(2011)}]{Haah11}%
  \BibitemOpen
  \bibfield  {author} {\bibinfo {author} {\bibfnamefont {J.}~\bibnamefont
  {Haah}},\ }\href {\doibase 10.1103/PhysRevA.83.042330} {\bibfield  {journal}
  {\bibinfo  {journal} {Phys. Rev. A}\ }\textbf {\bibinfo {volume} {83}},\
  \bibinfo {pages} {042330} (\bibinfo {year} {2011})}\BibitemShut {NoStop}%
\bibitem [{\citenamefont {Yoshida}(2013)}]{Yoshida13}%
  \BibitemOpen
  \bibfield  {author} {\bibinfo {author} {\bibfnamefont {B.}~\bibnamefont
  {Yoshida}},\ }\href {\doibase 10.1103/PhysRevB.88.125122} {\bibfield
  {journal} {\bibinfo  {journal} {Phys. Rev. B}\ }\textbf {\bibinfo {volume}
  {88}},\ \bibinfo {pages} {125122} (\bibinfo {year} {2013})}\BibitemShut
  {NoStop}%
\bibitem [{\citenamefont {Vijay}\ \emph {et~al.}(2015)\citenamefont {Vijay},
  \citenamefont {Haah},\ and\ \citenamefont {Fu}}]{Vijay15}%
  \BibitemOpen
  \bibfield  {author} {\bibinfo {author} {\bibfnamefont {S.}~\bibnamefont
  {Vijay}}, \bibinfo {author} {\bibfnamefont {J.}~\bibnamefont {Haah}}, \ and\
  \bibinfo {author} {\bibfnamefont {L.}~\bibnamefont {Fu}},\ }\href {\doibase
  10.1103/PhysRevB.92.235136} {\bibfield  {journal} {\bibinfo  {journal} {Phys.
  Rev. B}\ }\textbf {\bibinfo {volume} {92}},\ \bibinfo {pages} {235136}
  (\bibinfo {year} {2015})}\BibitemShut {NoStop}%
\bibitem [{\citenamefont {Williamson}(2016)}]{Williamson16}%
  \BibitemOpen
  \bibfield  {author} {\bibinfo {author} {\bibfnamefont {D.~J.}\ \bibnamefont
  {Williamson}},\ }\href {\doibase 10.1103/PhysRevB.94.155128} {\bibfield
  {journal} {\bibinfo  {journal} {Phys. Rev. B}\ }\textbf {\bibinfo {volume}
  {94}},\ \bibinfo {pages} {155128} (\bibinfo {year} {2016})}\BibitemShut
  {NoStop}%
\bibitem [{\citenamefont {Hal\'asz}\ \emph {et~al.}(2017)\citenamefont
  {Hal\'asz}, \citenamefont {Hsieh},\ and\ \citenamefont {Balents}}]{Halasz17}%
  \BibitemOpen
  \bibfield  {author} {\bibinfo {author} {\bibfnamefont {G.~B.}\ \bibnamefont
  {Hal\'asz}}, \bibinfo {author} {\bibfnamefont {T.~H.}\ \bibnamefont {Hsieh}},
  \ and\ \bibinfo {author} {\bibfnamefont {L.}~\bibnamefont {Balents}},\ }\href
  {\doibase 10.1103/PhysRevLett.119.257202} {\bibfield  {journal} {\bibinfo
  {journal} {Phys. Rev. Lett.}\ }\textbf {\bibinfo {volume} {119}},\ \bibinfo
  {pages} {257202} (\bibinfo {year} {2017})}\BibitemShut {NoStop}%
\bibitem [{\citenamefont {Ma}\ \emph {et~al.}(2017)\citenamefont {Ma},
  \citenamefont {Lake}, \citenamefont {Chen},\ and\ \citenamefont
  {Hermele}}]{HMa17}%
  \BibitemOpen
  \bibfield  {author} {\bibinfo {author} {\bibfnamefont {H.}~\bibnamefont
  {Ma}}, \bibinfo {author} {\bibfnamefont {E.}~\bibnamefont {Lake}}, \bibinfo
  {author} {\bibfnamefont {X.}~\bibnamefont {Chen}}, \ and\ \bibinfo {author}
  {\bibfnamefont {M.}~\bibnamefont {Hermele}},\ }\href {\doibase
  10.1103/PhysRevB.95.245126} {\bibfield  {journal} {\bibinfo  {journal} {Phys.
  Rev. B}\ }\textbf {\bibinfo {volume} {95}},\ \bibinfo {pages} {245126}
  (\bibinfo {year} {2017})}\BibitemShut {NoStop}%
\bibitem [{\citenamefont {Petrova}\ and\ \citenamefont
  {Regnault}(2017)}]{Petrova17}%
  \BibitemOpen
  \bibfield  {author} {\bibinfo {author} {\bibfnamefont {O.}~\bibnamefont
  {Petrova}}\ and\ \bibinfo {author} {\bibfnamefont {N.}~\bibnamefont
  {Regnault}},\ }\href {\doibase 10.1103/PhysRevB.96.224429} {\bibfield
  {journal} {\bibinfo  {journal} {Phys. Rev. B}\ }\textbf {\bibinfo {volume}
  {96}},\ \bibinfo {pages} {224429} (\bibinfo {year} {2017})}\BibitemShut
  {NoStop}%
\bibitem [{\citenamefont {Prem}\ \emph {et~al.}(2017)\citenamefont {Prem},
  \citenamefont {Haah},\ and\ \citenamefont {Nandkishore}}]{Prem17}%
  \BibitemOpen
  \bibfield  {author} {\bibinfo {author} {\bibfnamefont {A.}~\bibnamefont
  {Prem}}, \bibinfo {author} {\bibfnamefont {J.}~\bibnamefont {Haah}}, \ and\
  \bibinfo {author} {\bibfnamefont {R.}~\bibnamefont {Nandkishore}},\ }\href
  {\doibase 10.1103/PhysRevB.95.155133} {\bibfield  {journal} {\bibinfo
  {journal} {Phys. Rev. B}\ }\textbf {\bibinfo {volume} {95}},\ \bibinfo
  {pages} {155133} (\bibinfo {year} {2017})}\BibitemShut {NoStop}%
\bibitem [{\citenamefont {Pretko}(2017{\natexlab{a}})}]{Pretko17a}%
  \BibitemOpen
  \bibfield  {author} {\bibinfo {author} {\bibfnamefont {M.}~\bibnamefont
  {Pretko}},\ }\href {\doibase 10.1103/PhysRevB.95.115139} {\bibfield
  {journal} {\bibinfo  {journal} {Phys. Rev. B}\ }\textbf {\bibinfo {volume}
  {95}},\ \bibinfo {pages} {115139} (\bibinfo {year}
  {2017}{\natexlab{a}})}\BibitemShut {NoStop}%
\bibitem [{\citenamefont {Pretko}(2017{\natexlab{b}})}]{Pretko17b}%
  \BibitemOpen
  \bibfield  {author} {\bibinfo {author} {\bibfnamefont {M.}~\bibnamefont
  {Pretko}},\ }\href {\doibase 10.1103/PhysRevB.96.035119} {\bibfield
  {journal} {\bibinfo  {journal} {Phys. Rev. B}\ }\textbf {\bibinfo {volume}
  {96}},\ \bibinfo {pages} {035119} (\bibinfo {year}
  {2017}{\natexlab{b}})}\BibitemShut {NoStop}%
\bibitem [{\citenamefont {Pretko}(2017{\natexlab{c}})}]{Pretko17d}%
  \BibitemOpen
  \bibfield  {author} {\bibinfo {author} {\bibfnamefont {M.}~\bibnamefont
  {Pretko}},\ }\href {\doibase 10.1103/PhysRevB.96.125151} {\bibfield
  {journal} {\bibinfo  {journal} {Phys. Rev. B}\ }\textbf {\bibinfo {volume}
  {96}},\ \bibinfo {pages} {125151} (\bibinfo {year}
  {2017}{\natexlab{c}})}\BibitemShut {NoStop}%
\bibitem [{\citenamefont {Slagle}\ and\ \citenamefont
  {Kim}(2017{\natexlab{a}})}]{Slagle17a}%
  \BibitemOpen
  \bibfield  {author} {\bibinfo {author} {\bibfnamefont {K.}~\bibnamefont
  {Slagle}}\ and\ \bibinfo {author} {\bibfnamefont {Y.~B.}\ \bibnamefont
  {Kim}},\ }\href {\doibase 10.1103/PhysRevB.96.165106} {\bibfield  {journal}
  {\bibinfo  {journal} {Phys. Rev. B}\ }\textbf {\bibinfo {volume} {96}},\
  \bibinfo {pages} {165106} (\bibinfo {year} {2017}{\natexlab{a}})}\BibitemShut
  {NoStop}%
\bibitem [{\citenamefont {Slagle}\ and\ \citenamefont
  {Kim}(2017{\natexlab{b}})}]{Slagle17b}%
  \BibitemOpen
  \bibfield  {author} {\bibinfo {author} {\bibfnamefont {K.}~\bibnamefont
  {Slagle}}\ and\ \bibinfo {author} {\bibfnamefont {Y.~B.}\ \bibnamefont
  {Kim}},\ }\href {\doibase 10.1103/PhysRevB.96.195139} {\bibfield  {journal}
  {\bibinfo  {journal} {Phys. Rev. B}\ }\textbf {\bibinfo {volume} {96}},\
  \bibinfo {pages} {195139} (\bibinfo {year} {2017}{\natexlab{b}})}\BibitemShut
  {NoStop}%
\bibitem [{\citenamefont {Hsieh}\ and\ \citenamefont
  {Hal\'asz}(2017)}]{THHsieh17}%
  \BibitemOpen
  \bibfield  {author} {\bibinfo {author} {\bibfnamefont {T.~H.}\ \bibnamefont
  {Hsieh}}\ and\ \bibinfo {author} {\bibfnamefont {G.~B.}\ \bibnamefont
  {Hal\'asz}},\ }\href {\doibase 10.1103/PhysRevB.96.165105} {\bibfield
  {journal} {\bibinfo  {journal} {Phys. Rev. B}\ }\textbf {\bibinfo {volume}
  {96}},\ \bibinfo {pages} {165105} (\bibinfo {year} {2017})}\BibitemShut
  {NoStop}%
\bibitem [{\citenamefont {Vijay}()}]{Vijay17a}%
  \BibitemOpen
  \bibfield  {author} {\bibinfo {author} {\bibfnamefont {S.}~\bibnamefont
  {Vijay}},\ }\href {http://arxiv.org/abs/1701.00762} {\enquote {\bibinfo
  {title} {{Isotropic Layer Construction and Phase Diagram for Fracton
  Topological Phases}},}\ }\Eprint {http://arxiv.org/abs/arXiv:1701.00762}
  {arXiv:1701.00762} \BibitemShut {NoStop}%
\bibitem [{\citenamefont {Vijay}\ and\ \citenamefont {Fu}()}]{Vijay17b}%
  \BibitemOpen
  \bibfield  {author} {\bibinfo {author} {\bibfnamefont {S.}~\bibnamefont
  {Vijay}}\ and\ \bibinfo {author} {\bibfnamefont {L.}~\bibnamefont {Fu}},\
  }\href {http://arxiv.org/abs/1706.07070} {\enquote {\bibinfo {title} {{A
  Generalization of Non-Abelian Anyons in Three Dimensions}},}\ }\Eprint
  {http://arxiv.org/abs/arXiv:1706.07070} {arXiv:1706.07070} \BibitemShut
  {NoStop}%
\bibitem [{\citenamefont {Shi}\ and\ \citenamefont {Lu}(2018)}]{BShi18}%
  \BibitemOpen
  \bibfield  {author} {\bibinfo {author} {\bibfnamefont {B.}~\bibnamefont
  {Shi}}\ and\ \bibinfo {author} {\bibfnamefont {Y.-M.}\ \bibnamefont {Lu}},\
  }\href {\doibase 10.1103/PhysRevB.97.144106} {\bibfield  {journal} {\bibinfo
  {journal} {Phys. Rev. B}\ }\textbf {\bibinfo {volume} {97}},\ \bibinfo
  {pages} {144106} (\bibinfo {year} {2018})}\BibitemShut {NoStop}%
\bibitem [{\citenamefont {Bulmash}\ and\ \citenamefont
  {Barkeshli}({\natexlab{a}})}]{Bulmash18a}%
  \BibitemOpen
  \bibfield  {author} {\bibinfo {author} {\bibfnamefont {D.}~\bibnamefont
  {Bulmash}}\ and\ \bibinfo {author} {\bibfnamefont {M.}~\bibnamefont
  {Barkeshli}},\ }\href {http://arxiv.org/abs/1806.01855} {\enquote {\bibinfo
  {title} {{Generalized $U(1)$ Gauge Field Theories and Fractal Dynamics}},}\ }
  ({\natexlab{a}}),\ \Eprint {http://arxiv.org/abs/arXiv:1806.01855}
  {arXiv:1806.01855} \BibitemShut {NoStop}%
\bibitem [{\citenamefont {Bulmash}\ and\ \citenamefont
  {Barkeshli}(2018)}]{Bulmash18b}%
  \BibitemOpen
  \bibfield  {author} {\bibinfo {author} {\bibfnamefont {D.}~\bibnamefont
  {Bulmash}}\ and\ \bibinfo {author} {\bibfnamefont {M.}~\bibnamefont
  {Barkeshli}},\ }\href {\doibase 10.1103/PhysRevB.97.235112} {\bibfield
  {journal} {\bibinfo  {journal} {Phys. Rev. B}\ }\textbf {\bibinfo {volume}
  {97}},\ \bibinfo {pages} {235112} (\bibinfo {year} {2018})}\BibitemShut
  {NoStop}%
\bibitem [{\citenamefont {Devakul}\ \emph {et~al.}(2018)\citenamefont
  {Devakul}, \citenamefont {Parameswaran},\ and\ \citenamefont
  {Sondhi}}]{Devakul18b}%
  \BibitemOpen
  \bibfield  {author} {\bibinfo {author} {\bibfnamefont {T.}~\bibnamefont
  {Devakul}}, \bibinfo {author} {\bibfnamefont {S.~A.}\ \bibnamefont
  {Parameswaran}}, \ and\ \bibinfo {author} {\bibfnamefont {S.~L.}\
  \bibnamefont {Sondhi}},\ }\href {\doibase 10.1103/PhysRevB.97.041110}
  {\bibfield  {journal} {\bibinfo  {journal} {Phys. Rev. B}\ }\textbf {\bibinfo
  {volume} {97}},\ \bibinfo {pages} {041110} (\bibinfo {year}
  {2018})}\BibitemShut {NoStop}%
\bibitem [{\citenamefont {Gromov}()}]{Gromov18}%
  \BibitemOpen
  \bibfield  {author} {\bibinfo {author} {\bibfnamefont {A.}~\bibnamefont
  {Gromov}},\ }\href {http://arxiv.org/abs/1812.05104} {\enquote {\bibinfo
  {title} {{Towards classification of Fracton phases: the multipole
  algebra}},}\ }\Eprint {http://arxiv.org/abs/arXiv:1812.05104}
  {arXiv:1812.05104} \BibitemShut {NoStop}%
\bibitem [{\citenamefont {Ma}\ \emph {et~al.}(2018{\natexlab{a}})\citenamefont
  {Ma}, \citenamefont {Schmitz}, \citenamefont {Parameswaran}, \citenamefont
  {Hermele},\ and\ \citenamefont {Nandkishore}}]{HMa18a}%
  \BibitemOpen
  \bibfield  {author} {\bibinfo {author} {\bibfnamefont {H.}~\bibnamefont
  {Ma}}, \bibinfo {author} {\bibfnamefont {A.~T.}\ \bibnamefont {Schmitz}},
  \bibinfo {author} {\bibfnamefont {S.~A.}\ \bibnamefont {Parameswaran}},
  \bibinfo {author} {\bibfnamefont {M.}~\bibnamefont {Hermele}}, \ and\
  \bibinfo {author} {\bibfnamefont {R.~M.}\ \bibnamefont {Nandkishore}},\
  }\href {\doibase 10.1103/PhysRevB.97.125101} {\bibfield  {journal} {\bibinfo
  {journal} {Phys. Rev. B}\ }\textbf {\bibinfo {volume} {97}},\ \bibinfo
  {pages} {125101} (\bibinfo {year} {2018}{\natexlab{a}})}\BibitemShut
  {NoStop}%
\bibitem [{\citenamefont {Ma}\ \emph {et~al.}(2018{\natexlab{b}})\citenamefont
  {Ma}, \citenamefont {Hermele},\ and\ \citenamefont {Chen}}]{HMa18b}%
  \BibitemOpen
  \bibfield  {author} {\bibinfo {author} {\bibfnamefont {H.}~\bibnamefont
  {Ma}}, \bibinfo {author} {\bibfnamefont {M.}~\bibnamefont {Hermele}}, \ and\
  \bibinfo {author} {\bibfnamefont {X.}~\bibnamefont {Chen}},\ }\href {\doibase
  10.1103/PhysRevB.98.035111} {\bibfield  {journal} {\bibinfo  {journal} {Phys.
  Rev. B}\ }\textbf {\bibinfo {volume} {98}},\ \bibinfo {pages} {035111}
  (\bibinfo {year} {2018}{\natexlab{b}})}\BibitemShut {NoStop}%
\bibitem [{\citenamefont {He}\ \emph {et~al.}(2018)\citenamefont {He},
  \citenamefont {Zheng}, \citenamefont {Bernevig},\ and\ \citenamefont
  {Regnault}}]{HHe18}%
  \BibitemOpen
  \bibfield  {author} {\bibinfo {author} {\bibfnamefont {H.}~\bibnamefont
  {He}}, \bibinfo {author} {\bibfnamefont {Y.}~\bibnamefont {Zheng}}, \bibinfo
  {author} {\bibfnamefont {B.~A.}\ \bibnamefont {Bernevig}}, \ and\ \bibinfo
  {author} {\bibfnamefont {N.}~\bibnamefont {Regnault}},\ }\href {\doibase
  10.1103/PhysRevB.97.125102} {\bibfield  {journal} {\bibinfo  {journal} {Phys.
  Rev. B}\ }\textbf {\bibinfo {volume} {97}},\ \bibinfo {pages} {125102}
  (\bibinfo {year} {2018})}\BibitemShut {NoStop}%
\bibitem [{\citenamefont {Pai}\ and\ \citenamefont {Pretko}(2018)}]{Pai18}%
  \BibitemOpen
  \bibfield  {author} {\bibinfo {author} {\bibfnamefont {S.}~\bibnamefont
  {Pai}}\ and\ \bibinfo {author} {\bibfnamefont {M.}~\bibnamefont {Pretko}},\
  }\href {\doibase 10.1103/PhysRevB.97.235102} {\bibfield  {journal} {\bibinfo
  {journal} {Phys. Rev. B}\ }\textbf {\bibinfo {volume} {97}},\ \bibinfo
  {pages} {235102} (\bibinfo {year} {2018})}\BibitemShut {NoStop}%
\bibitem [{\citenamefont {Prem}\ \emph
  {et~al.}(2018{\natexlab{a}})\citenamefont {Prem}, \citenamefont {Pretko},\
  and\ \citenamefont {Nandkishore}}]{Prem18a}%
  \BibitemOpen
  \bibfield  {author} {\bibinfo {author} {\bibfnamefont {A.}~\bibnamefont
  {Prem}}, \bibinfo {author} {\bibfnamefont {M.}~\bibnamefont {Pretko}}, \ and\
  \bibinfo {author} {\bibfnamefont {R.~M.}\ \bibnamefont {Nandkishore}},\
  }\href {\doibase 10.1103/PhysRevB.97.085116} {\bibfield  {journal} {\bibinfo
  {journal} {Phys. Rev. B}\ }\textbf {\bibinfo {volume} {97}},\ \bibinfo
  {pages} {085116} (\bibinfo {year} {2018}{\natexlab{a}})}\BibitemShut
  {NoStop}%
\bibitem [{\citenamefont {Prem}\ \emph
  {et~al.}(2018{\natexlab{b}})\citenamefont {Prem}, \citenamefont {Vijay},
  \citenamefont {Chou}, \citenamefont {Pretko},\ and\ \citenamefont
  {Nandkishore}}]{Prem18b}%
  \BibitemOpen
  \bibfield  {author} {\bibinfo {author} {\bibfnamefont {A.}~\bibnamefont
  {Prem}}, \bibinfo {author} {\bibfnamefont {S.}~\bibnamefont {Vijay}},
  \bibinfo {author} {\bibfnamefont {Y.-Z.}\ \bibnamefont {Chou}}, \bibinfo
  {author} {\bibfnamefont {M.}~\bibnamefont {Pretko}}, \ and\ \bibinfo {author}
  {\bibfnamefont {R.~M.}\ \bibnamefont {Nandkishore}},\ }\href {\doibase
  10.1103/PhysRevB.98.165140} {\bibfield  {journal} {\bibinfo  {journal} {Phys.
  Rev. B}\ }\textbf {\bibinfo {volume} {98}},\ \bibinfo {pages} {165140}
  (\bibinfo {year} {2018}{\natexlab{b}})}\BibitemShut {NoStop}%
\bibitem [{\citenamefont {Pretko}\ and\ \citenamefont
  {Radzihovsky}(2018{\natexlab{a}})}]{Pretko18b}%
  \BibitemOpen
  \bibfield  {author} {\bibinfo {author} {\bibfnamefont {M.}~\bibnamefont
  {Pretko}}\ and\ \bibinfo {author} {\bibfnamefont {L.}~\bibnamefont
  {Radzihovsky}},\ }\href {\doibase 10.1103/PhysRevLett.120.195301} {\bibfield
  {journal} {\bibinfo  {journal} {Phys. Rev. Lett.}\ }\textbf {\bibinfo
  {volume} {120}},\ \bibinfo {pages} {195301} (\bibinfo {year}
  {2018}{\natexlab{a}})}\BibitemShut {NoStop}%
\bibitem [{\citenamefont {Pretko}\ and\ \citenamefont
  {Radzihovsky}(2018{\natexlab{b}})}]{Pretko18c}%
  \BibitemOpen
  \bibfield  {author} {\bibinfo {author} {\bibfnamefont {M.}~\bibnamefont
  {Pretko}}\ and\ \bibinfo {author} {\bibfnamefont {L.}~\bibnamefont
  {Radzihovsky}},\ }\href {\doibase 10.1103/PhysRevLett.121.235301} {\bibfield
  {journal} {\bibinfo  {journal} {Phys. Rev. Lett.}\ }\textbf {\bibinfo
  {volume} {121}},\ \bibinfo {pages} {235301} (\bibinfo {year}
  {2018}{\natexlab{b}})}\BibitemShut {NoStop}%
\bibitem [{\citenamefont {Schmitz}\ \emph {et~al.}(2018)\citenamefont
  {Schmitz}, \citenamefont {Ma}, \citenamefont {Nandkishore},\ and\
  \citenamefont {Parameswaran}}]{Schmitz18b}%
  \BibitemOpen
  \bibfield  {author} {\bibinfo {author} {\bibfnamefont {A.~T.}\ \bibnamefont
  {Schmitz}}, \bibinfo {author} {\bibfnamefont {H.}~\bibnamefont {Ma}},
  \bibinfo {author} {\bibfnamefont {R.~M.}\ \bibnamefont {Nandkishore}}, \ and\
  \bibinfo {author} {\bibfnamefont {S.~A.}\ \bibnamefont {Parameswaran}},\
  }\href {\doibase 10.1103/PhysRevB.97.134426} {\bibfield  {journal} {\bibinfo
  {journal} {Phys. Rev. B}\ }\textbf {\bibinfo {volume} {97}},\ \bibinfo
  {pages} {134426} (\bibinfo {year} {2018})}\BibitemShut {NoStop}%
\bibitem [{\citenamefont {Shirley}\ \emph {et~al.}({\natexlab{a}})\citenamefont
  {Shirley}, \citenamefont {Slagle},\ and\ \citenamefont {Chen}}]{Shirley18a}%
  \BibitemOpen
  \bibfield  {author} {\bibinfo {author} {\bibfnamefont {W.}~\bibnamefont
  {Shirley}}, \bibinfo {author} {\bibfnamefont {K.}~\bibnamefont {Slagle}}, \
  and\ \bibinfo {author} {\bibfnamefont {X.}~\bibnamefont {Chen}},\ }\href
  {http://arxiv.org/abs/1806.08625} {\enquote {\bibinfo {title} {{Fractional
  excitations in foliated fracton phases}},}\ } ({\natexlab{a}}),\ \Eprint
  {http://arxiv.org/abs/arXiv:1806.08625} {arXiv:1806.08625} \BibitemShut
  {NoStop}%
\bibitem [{\citenamefont {Shirley}\ \emph {et~al.}(2018)\citenamefont
  {Shirley}, \citenamefont {Slagle}, \citenamefont {Wang},\ and\ \citenamefont
  {Chen}}]{Shirley18b}%
  \BibitemOpen
  \bibfield  {author} {\bibinfo {author} {\bibfnamefont {W.}~\bibnamefont
  {Shirley}}, \bibinfo {author} {\bibfnamefont {K.}~\bibnamefont {Slagle}},
  \bibinfo {author} {\bibfnamefont {Z.}~\bibnamefont {Wang}}, \ and\ \bibinfo
  {author} {\bibfnamefont {X.}~\bibnamefont {Chen}},\ }\href {\doibase
  10.1103/PhysRevX.8.031051} {\bibfield  {journal} {\bibinfo  {journal} {Phys.
  Rev. X}\ }\textbf {\bibinfo {volume} {8}},\ \bibinfo {pages} {031051}
  (\bibinfo {year} {2018})}\BibitemShut {NoStop}%
\bibitem [{\citenamefont {Slagle}\ and\ \citenamefont {Kim}(2018)}]{Slagle18b}%
  \BibitemOpen
  \bibfield  {author} {\bibinfo {author} {\bibfnamefont {K.}~\bibnamefont
  {Slagle}}\ and\ \bibinfo {author} {\bibfnamefont {Y.~B.}\ \bibnamefont
  {Kim}},\ }\href {\doibase 10.1103/PhysRevB.97.165106} {\bibfield  {journal}
  {\bibinfo  {journal} {Phys. Rev. B}\ }\textbf {\bibinfo {volume} {97}},\
  \bibinfo {pages} {165106} (\bibinfo {year} {2018})}\BibitemShut {NoStop}%
\bibitem [{\citenamefont {Williamson}\ \emph {et~al.}()\citenamefont
  {Williamson}, \citenamefont {Bi},\ and\ \citenamefont
  {Cheng}}]{Williamson18}%
  \BibitemOpen
  \bibfield  {author} {\bibinfo {author} {\bibfnamefont {D.~J.}\ \bibnamefont
  {Williamson}}, \bibinfo {author} {\bibfnamefont {Z.}~\bibnamefont {Bi}}, \
  and\ \bibinfo {author} {\bibfnamefont {M.}~\bibnamefont {Cheng}},\ }\href
  {http://arxiv.org/abs/1809.10275} {\enquote {\bibinfo {title} {{Fractonic
  Matter in Symmetry-Enriched U(1) Gauge Theory}},}\ }\Eprint
  {http://arxiv.org/abs/arXiv:1809.10275} {arXiv:1809.10275} \BibitemShut
  {NoStop}%
\bibitem [{\citenamefont {You}\ \emph {et~al.}({\natexlab{a}})\citenamefont
  {You}, \citenamefont {Devakul}, \citenamefont {Burnell},\ and\ \citenamefont
  {Sondhi}}]{YYou18a}%
  \BibitemOpen
  \bibfield  {author} {\bibinfo {author} {\bibfnamefont {Y.}~\bibnamefont
  {You}}, \bibinfo {author} {\bibfnamefont {T.}~\bibnamefont {Devakul}},
  \bibinfo {author} {\bibfnamefont {F.~J.}\ \bibnamefont {Burnell}}, \ and\
  \bibinfo {author} {\bibfnamefont {S.~L.}\ \bibnamefont {Sondhi}},\ }\href
  {http://arxiv.org/abs/1805.09800} {\enquote {\bibinfo {title} {{Symmetric
  Fracton Matter: Twisted and Enriched}},}\ } ({\natexlab{a}}),\ \Eprint
  {http://arxiv.org/abs/arXiv:1805.09800} {arXiv:1805.09800} \BibitemShut
  {NoStop}%
\bibitem [{\citenamefont {You}\ \emph {et~al.}({\natexlab{b}})\citenamefont
  {You}, \citenamefont {Litinski},\ and\ \citenamefont {{von
  Oppen}}}]{YYou18b}%
  \BibitemOpen
  \bibfield  {author} {\bibinfo {author} {\bibfnamefont {Y.}~\bibnamefont
  {You}}, \bibinfo {author} {\bibfnamefont {D.}~\bibnamefont {Litinski}}, \
  and\ \bibinfo {author} {\bibfnamefont {F.}~\bibnamefont {{von Oppen}}},\
  }\href {http://arxiv.org/abs/1810.10556} {\enquote {\bibinfo {title} {{Higher
  order topological superconductors as generators of quantum codes}},}\ }
  ({\natexlab{b}}),\ \Eprint {http://arxiv.org/abs/arXiv:1810.10556}
  {arXiv:1810.10556} \BibitemShut {NoStop}%
\bibitem [{\citenamefont {You}\ and\ \citenamefont {{von Oppen}}()}]{YYou18c}%
  \BibitemOpen
  \bibfield  {author} {\bibinfo {author} {\bibfnamefont {Y.}~\bibnamefont
  {You}}\ and\ \bibinfo {author} {\bibfnamefont {F.}~\bibnamefont {{von
  Oppen}}},\ }\href {http://arxiv.org/abs/1812.06091} {\enquote {\bibinfo
  {title} {{Majorana Quantum Lego, a Route Towards Fracton Matter}},}\ }\Eprint
  {http://arxiv.org/abs/arXiv:1812.06091} {arXiv:1812.06091} \BibitemShut
  {NoStop}%
\bibitem [{\citenamefont {Bulmash}\ and\ \citenamefont
  {Iadecola}(2019)}]{Bulmash19a}%
  \BibitemOpen
  \bibfield  {author} {\bibinfo {author} {\bibfnamefont {D.}~\bibnamefont
  {Bulmash}}\ and\ \bibinfo {author} {\bibfnamefont {T.}~\bibnamefont
  {Iadecola}},\ }\href {\doibase 10.1103/PhysRevB.99.125132} {\bibfield
  {journal} {\bibinfo  {journal} {Phys. Rev. B}\ }\textbf {\bibinfo {volume}
  {99}},\ \bibinfo {pages} {125132} (\bibinfo {year} {2019})}\BibitemShut
  {NoStop}%
\bibitem [{\citenamefont {Bulmash}\ and\ \citenamefont
  {Barkeshli}({\natexlab{b}})}]{Bulmash19b}%
  \BibitemOpen
  \bibfield  {author} {\bibinfo {author} {\bibfnamefont {D.}~\bibnamefont
  {Bulmash}}\ and\ \bibinfo {author} {\bibfnamefont {M.}~\bibnamefont
  {Barkeshli}},\ }\href {http://arxiv.org/abs/1905.05771} {\enquote {\bibinfo
  {title} {{Gauging fractons: immobile non-Abelian quasiparticles, fractals,
  and position-dependent degeneracies}},}\ } ({\natexlab{b}}),\ \Eprint
  {http://arxiv.org/abs/arXiv:1905.05771} {arXiv:1905.05771} \BibitemShut
  {NoStop}%
\bibitem [{\citenamefont {Dua}\ \emph {et~al.}(2019)\citenamefont {Dua},
  \citenamefont {Williamson}, \citenamefont {Haah},\ and\ \citenamefont
  {Cheng}}]{Dua19}%
  \BibitemOpen
  \bibfield  {author} {\bibinfo {author} {\bibfnamefont {A.}~\bibnamefont
  {Dua}}, \bibinfo {author} {\bibfnamefont {D.~J.}\ \bibnamefont {Williamson}},
  \bibinfo {author} {\bibfnamefont {J.}~\bibnamefont {Haah}}, \ and\ \bibinfo
  {author} {\bibfnamefont {M.}~\bibnamefont {Cheng}},\ }\href {\doibase
  10.1103/PhysRevB.99.245135} {\bibfield  {journal} {\bibinfo  {journal} {Phys.
  Rev. B}\ }\textbf {\bibinfo {volume} {99}},\ \bibinfo {pages} {245135}
  (\bibinfo {year} {2019})}\BibitemShut {NoStop}%
\bibitem [{\citenamefont {Gromov}(2019)}]{Gromov19}%
  \BibitemOpen
  \bibfield  {author} {\bibinfo {author} {\bibfnamefont {A.}~\bibnamefont
  {Gromov}},\ }\href {\doibase 10.1103/PhysRevLett.122.076403} {\bibfield
  {journal} {\bibinfo  {journal} {Phys. Rev. Lett.}\ }\textbf {\bibinfo
  {volume} {122}},\ \bibinfo {pages} {076403} (\bibinfo {year}
  {2019})}\BibitemShut {NoStop}%
\bibitem [{\citenamefont {Yan}(2019)}]{HYan19a}%
  \BibitemOpen
  \bibfield  {author} {\bibinfo {author} {\bibfnamefont {H.}~\bibnamefont
  {Yan}},\ }\href {\doibase 10.1103/PhysRevB.99.155126} {\bibfield  {journal}
  {\bibinfo  {journal} {Phys. Rev. B}\ }\textbf {\bibinfo {volume} {99}},\
  \bibinfo {pages} {155126} (\bibinfo {year} {2019})}\BibitemShut {NoStop}%
\bibitem [{\citenamefont {Yan}()}]{HYan19b}%
  \BibitemOpen
  \bibfield  {author} {\bibinfo {author} {\bibfnamefont {H.}~\bibnamefont
  {Yan}},\ }\href {http://arxiv.org/abs/1906.02305} {\enquote {\bibinfo {title}
  {{Hyperbolic Fracton Model, Subsystem Symmetry, and Holography II: The Dual
  Eight-Vertex Model}},}\ }\Eprint {http://arxiv.org/abs/arXiv:1906.02305}
  {arXiv:1906.02305} \BibitemShut {NoStop}%
\bibitem [{\citenamefont {Song}\ \emph {et~al.}(2019)\citenamefont {Song},
  \citenamefont {Prem}, \citenamefont {Huang},\ and\ \citenamefont
  {Martin-Delgado}}]{HSong19}%
  \BibitemOpen
  \bibfield  {author} {\bibinfo {author} {\bibfnamefont {H.}~\bibnamefont
  {Song}}, \bibinfo {author} {\bibfnamefont {A.}~\bibnamefont {Prem}}, \bibinfo
  {author} {\bibfnamefont {S.-J.}\ \bibnamefont {Huang}}, \ and\ \bibinfo
  {author} {\bibfnamefont {M.~A.}\ \bibnamefont {Martin-Delgado}},\ }\href
  {\doibase 10.1103/PhysRevB.99.155118} {\bibfield  {journal} {\bibinfo
  {journal} {Phys. Rev. B}\ }\textbf {\bibinfo {volume} {99}},\ \bibinfo
  {pages} {155118} (\bibinfo {year} {2019})}\BibitemShut {NoStop}%
\bibitem [{\citenamefont {Tian}\ and\ \citenamefont {Wang}()}]{Tian19}%
  \BibitemOpen
  \bibfield  {author} {\bibinfo {author} {\bibfnamefont {K.~T.}\ \bibnamefont
  {Tian}}\ and\ \bibinfo {author} {\bibfnamefont {Z.}~\bibnamefont {Wang}},\
  }\href {http://arxiv.org/abs/1902.04543} {\enquote {\bibinfo {title}
  {{Generalized Haah Codes and Fracton Models}},}\ }\Eprint
  {http://arxiv.org/abs/arXiv:1902.04543} {arXiv:1902.04543} \BibitemShut
  {NoStop}%
\bibitem [{\citenamefont {Pai}\ and\ \citenamefont {Hermele}()}]{Pai19}%
  \BibitemOpen
  \bibfield  {author} {\bibinfo {author} {\bibfnamefont {S.}~\bibnamefont
  {Pai}}\ and\ \bibinfo {author} {\bibfnamefont {M.}~\bibnamefont {Hermele}},\
  }\href {http://arxiv.org/abs/1903.11625} {\enquote {\bibinfo {title}
  {{Fracton fusion and statistics}},}\ }\Eprint
  {http://arxiv.org/abs/arXiv:1903.11625} {arXiv:1903.11625} \BibitemShut
  {NoStop}%
\bibitem [{\citenamefont {Prem}\ \emph {et~al.}(2019)\citenamefont {Prem},
  \citenamefont {Huang}, \citenamefont {Song},\ and\ \citenamefont
  {Hermele}}]{Prem19a}%
  \BibitemOpen
  \bibfield  {author} {\bibinfo {author} {\bibfnamefont {A.}~\bibnamefont
  {Prem}}, \bibinfo {author} {\bibfnamefont {S.-J.}\ \bibnamefont {Huang}},
  \bibinfo {author} {\bibfnamefont {H.}~\bibnamefont {Song}}, \ and\ \bibinfo
  {author} {\bibfnamefont {M.}~\bibnamefont {Hermele}},\ }\href {\doibase
  10.1103/PhysRevX.9.021010} {\bibfield  {journal} {\bibinfo  {journal} {Phys.
  Rev. X}\ }\textbf {\bibinfo {volume} {9}},\ \bibinfo {pages} {021010}
  (\bibinfo {year} {2019})}\BibitemShut {NoStop}%
\bibitem [{\citenamefont {Prem}\ and\ \citenamefont {Williamson}()}]{Prem19b}%
  \BibitemOpen
  \bibfield  {author} {\bibinfo {author} {\bibfnamefont {A.}~\bibnamefont
  {Prem}}\ and\ \bibinfo {author} {\bibfnamefont {D.~J.}\ \bibnamefont
  {Williamson}},\ }\href {http://arxiv.org/abs/1905.06309} {\enquote {\bibinfo
  {title} {{Gauging permutation symmetries as a route to non-Abelian
  fractons}},}\ }\Eprint {http://arxiv.org/abs/arXiv:1905.06309}
  {arXiv:1905.06309} \BibitemShut {NoStop}%
\bibitem [{\citenamefont {Shirley}\ \emph
  {et~al.}(2019{\natexlab{a}})\citenamefont {Shirley}, \citenamefont {Slagle},\
  and\ \citenamefont {Chen}}]{Shirley19a}%
  \BibitemOpen
  \bibfield  {author} {\bibinfo {author} {\bibfnamefont {W.}~\bibnamefont
  {Shirley}}, \bibinfo {author} {\bibfnamefont {K.}~\bibnamefont {Slagle}}, \
  and\ \bibinfo {author} {\bibfnamefont {X.}~\bibnamefont {Chen}},\ }\href
  {\doibase 10.1103/PhysRevB.99.115123} {\bibfield  {journal} {\bibinfo
  {journal} {Phys. Rev. B}\ }\textbf {\bibinfo {volume} {99}},\ \bibinfo
  {pages} {115123} (\bibinfo {year} {2019}{\natexlab{a}})}\BibitemShut
  {NoStop}%
\bibitem [{\citenamefont {Shirley}\ \emph
  {et~al.}(2019{\natexlab{b}})\citenamefont {Shirley}, \citenamefont {Slagle},\
  and\ \citenamefont {Chen}}]{Shirley19c}%
  \BibitemOpen
  \bibfield  {author} {\bibinfo {author} {\bibfnamefont {W.}~\bibnamefont
  {Shirley}}, \bibinfo {author} {\bibfnamefont {K.}~\bibnamefont {Slagle}}, \
  and\ \bibinfo {author} {\bibfnamefont {X.}~\bibnamefont {Chen}},\ }\href
  {\doibase 10.21468/SciPostPhys.6.4.041} {\bibfield  {journal} {\bibinfo
  {journal} {SciPost Phys.}\ }\textbf {\bibinfo {volume} {6}},\ \bibinfo
  {pages} {41} (\bibinfo {year} {2019}{\natexlab{b}})}\BibitemShut {NoStop}%
\bibitem [{\citenamefont {Shirley}\ \emph {et~al.}({\natexlab{b}})\citenamefont
  {Shirley}, \citenamefont {Slagle},\ and\ \citenamefont {Chen}}]{Shirley19d}%
  \BibitemOpen
  \bibfield  {author} {\bibinfo {author} {\bibfnamefont {W.}~\bibnamefont
  {Shirley}}, \bibinfo {author} {\bibfnamefont {K.}~\bibnamefont {Slagle}}, \
  and\ \bibinfo {author} {\bibfnamefont {X.}~\bibnamefont {Chen}},\ }\href
  {http://arxiv.org/abs/1907.09048} {\enquote {\bibinfo {title} {{Twisted
  foliated fracton phases}},}\ } ({\natexlab{b}}),\ \Eprint
  {http://arxiv.org/abs/arXiv:1907.09048} {arXiv:1907.09048} \BibitemShut
  {NoStop}%
\bibitem [{\citenamefont {Slagle}\ \emph {et~al.}(2019)\citenamefont {Slagle},
  \citenamefont {Aasen},\ and\ \citenamefont {Williamson}}]{Slagle19}%
  \BibitemOpen
  \bibfield  {author} {\bibinfo {author} {\bibfnamefont {K.}~\bibnamefont
  {Slagle}}, \bibinfo {author} {\bibfnamefont {D.}~\bibnamefont {Aasen}}, \
  and\ \bibinfo {author} {\bibfnamefont {D.}~\bibnamefont {Williamson}},\
  }\href {\doibase 10.21468/SciPostPhys.6.4.043} {\bibfield  {journal}
  {\bibinfo  {journal} {SciPost Phys.}\ }\textbf {\bibinfo {volume} {6}},\
  \bibinfo {pages} {43} (\bibinfo {year} {2019})}\BibitemShut {NoStop}%
\bibitem [{\citenamefont {Sous}\ and\ \citenamefont {Pretko}()}]{Sous19}%
  \BibitemOpen
  \bibfield  {author} {\bibinfo {author} {\bibfnamefont {J.}~\bibnamefont
  {Sous}}\ and\ \bibinfo {author} {\bibfnamefont {M.}~\bibnamefont {Pretko}},\
  }\href {http://arxiv.org/abs/1904.08424} {\enquote {\bibinfo {title}
  {{Fractons from Polarons and Hole-Doped Antiferromagnets: Microscopic Models
  and Realization}},}\ }\Eprint {http://arxiv.org/abs/arXiv:1904.08424}
  {arXiv:1904.08424} \BibitemShut {NoStop}%
\bibitem [{\citenamefont {Wang}\ \emph {et~al.}()\citenamefont {Wang},
  \citenamefont {Shirley},\ and\ \citenamefont {Chen}}]{TWang19}%
  \BibitemOpen
  \bibfield  {author} {\bibinfo {author} {\bibfnamefont {T.}~\bibnamefont
  {Wang}}, \bibinfo {author} {\bibfnamefont {W.}~\bibnamefont {Shirley}}, \
  and\ \bibinfo {author} {\bibfnamefont {X.}~\bibnamefont {Chen}},\ }\href
  {http://arxiv.org/abs/1904.01111} {\enquote {\bibinfo {title} {{Foliated
  fracton order in the Majorana checkerboard model}},}\ }\Eprint
  {http://arxiv.org/abs/arXiv:1904.01111} {arXiv:1904.01111} \BibitemShut
  {NoStop}%
\bibitem [{\citenamefont {You}\ \emph {et~al.}({\natexlab{c}})\citenamefont
  {You}, \citenamefont {Devakul}, \citenamefont {Sondhi},\ and\ \citenamefont
  {Burnell}}]{YYou19b}%
  \BibitemOpen
  \bibfield  {author} {\bibinfo {author} {\bibfnamefont {Y.}~\bibnamefont
  {You}}, \bibinfo {author} {\bibfnamefont {T.}~\bibnamefont {Devakul}},
  \bibinfo {author} {\bibfnamefont {S.~L.}\ \bibnamefont {Sondhi}}, \ and\
  \bibinfo {author} {\bibfnamefont {F.~J.}\ \bibnamefont {Burnell}},\ }\href
  {http://arxiv.org/abs/1904.11530} {\enquote {\bibinfo {title} {{Fractonic
  Chern-Simons and BF theories}},}\ } ({\natexlab{c}}),\ \Eprint
  {http://arxiv.org/abs/arXiv:1904.11530} {arXiv:1904.11530} \BibitemShut
  {NoStop}%
\bibitem [{\citenamefont {Nandkishore}\ and\ \citenamefont
  {Hermele}(2019)}]{Nandkishore19}%
  \BibitemOpen
  \bibfield  {author} {\bibinfo {author} {\bibfnamefont {R.~M.}\ \bibnamefont
  {Nandkishore}}\ and\ \bibinfo {author} {\bibfnamefont {M.}~\bibnamefont
  {Hermele}},\ }\href {\doibase 10.1146/annurev-conmatphys-031218-013604}
  {\bibfield  {journal} {\bibinfo  {journal} {Ann. Rev. Condens. Matter Phys.}\
  }\textbf {\bibinfo {volume} {10}},\ \bibinfo {pages} {295} (\bibinfo {year}
  {2019})}\BibitemShut {NoStop}%
\bibitem [{Note1()}]{Note1}%
  \BibitemOpen
  \bibinfo {note} {As a remark, while our models do not possess ``fractons'' as
  strictly immobile excitations, we abuse ``fracton models'' or ``fracton
  topological orders'' to emphasize that the corresponding models are still
  distinguished from the conventional topological orders or their decoulpled
  stacks.}\BibitemShut {Stop}%
\bibitem [{\citenamefont {Kitaev}(2003)}]{Kitaev03}%
  \BibitemOpen
  \bibfield  {author} {\bibinfo {author} {\bibfnamefont {A.}~\bibnamefont
  {Kitaev}},\ }\href {\doibase 10.1016/S0003-4916(02)00018-0} {\bibfield
  {journal} {\bibinfo  {journal} {Ann. Phys.}\ }\textbf {\bibinfo {volume}
  {303}},\ \bibinfo {pages} {2 } (\bibinfo {year} {2003})}\BibitemShut
  {NoStop}%
\bibitem [{\citenamefont {Bais}\ and\ \citenamefont
  {Slingerland}(2009)}]{Bais09}%
  \BibitemOpen
  \bibfield  {author} {\bibinfo {author} {\bibfnamefont {F.~A.}\ \bibnamefont
  {Bais}}\ and\ \bibinfo {author} {\bibfnamefont {J.~K.}\ \bibnamefont
  {Slingerland}},\ }\href {\doibase 10.1103/PhysRevB.79.045316} {\bibfield
  {journal} {\bibinfo  {journal} {Phys. Rev. B}\ }\textbf {\bibinfo {volume}
  {79}},\ \bibinfo {pages} {045316} (\bibinfo {year} {2009})}\BibitemShut
  {NoStop}%
\bibitem [{\citenamefont {Eli\"ens}\ \emph {et~al.}(2014)\citenamefont
  {Eli\"ens}, \citenamefont {Romers},\ and\ \citenamefont {Bais}}]{Eliens14}%
  \BibitemOpen
  \bibfield  {author} {\bibinfo {author} {\bibfnamefont {I.~S.}\ \bibnamefont
  {Eli\"ens}}, \bibinfo {author} {\bibfnamefont {J.~C.}\ \bibnamefont
  {Romers}}, \ and\ \bibinfo {author} {\bibfnamefont {F.~A.}\ \bibnamefont
  {Bais}},\ }\href {\doibase 10.1103/PhysRevB.90.195130} {\bibfield  {journal}
  {\bibinfo  {journal} {Phys. Rev. B}\ }\textbf {\bibinfo {volume} {90}},\
  \bibinfo {pages} {195130} (\bibinfo {year} {2014})}\BibitemShut {NoStop}%
\bibitem [{\citenamefont {Kong}(2014)}]{LKong14}%
  \BibitemOpen
  \bibfield  {author} {\bibinfo {author} {\bibfnamefont {L.}~\bibnamefont
  {Kong}},\ }\href {\doibase 10.1016/j.nuclphysb.2014.07.003} {\bibfield
  {journal} {\bibinfo  {journal} {Nucl. Phys. B}\ }\textbf {\bibinfo {volume}
  {886}},\ \bibinfo {pages} {436 } (\bibinfo {year} {2014})}\BibitemShut
  {NoStop}%
\bibitem [{\citenamefont {Neupert}\ \emph {et~al.}(2016)\citenamefont
  {Neupert}, \citenamefont {He}, \citenamefont {von Keyserlingk}, \citenamefont
  {Sierra},\ and\ \citenamefont {Bernevig}}]{Neupert16}%
  \BibitemOpen
  \bibfield  {author} {\bibinfo {author} {\bibfnamefont {T.}~\bibnamefont
  {Neupert}}, \bibinfo {author} {\bibfnamefont {H.}~\bibnamefont {He}},
  \bibinfo {author} {\bibfnamefont {C.}~\bibnamefont {von Keyserlingk}},
  \bibinfo {author} {\bibfnamefont {G.}~\bibnamefont {Sierra}}, \ and\ \bibinfo
  {author} {\bibfnamefont {B.~A.}\ \bibnamefont {Bernevig}},\ }\href {\doibase
  10.1103/PhysRevB.93.115103} {\bibfield  {journal} {\bibinfo  {journal} {Phys.
  Rev. B}\ }\textbf {\bibinfo {volume} {93}},\ \bibinfo {pages} {115103}
  (\bibinfo {year} {2016})}\BibitemShut {NoStop}%
\bibitem [{\citenamefont {Burnell}(2018)}]{Burnell18}%
  \BibitemOpen
  \bibfield  {author} {\bibinfo {author} {\bibfnamefont {F.}~\bibnamefont
  {Burnell}},\ }\href {\doibase 10.1146/annurev-conmatphys-033117-054154}
  {\bibfield  {journal} {\bibinfo  {journal} {Ann. Rev. Condens. Matter Phys.}\
  }\textbf {\bibinfo {volume} {9}},\ \bibinfo {pages} {307} (\bibinfo {year}
  {2018})}\BibitemShut {NoStop}%
\bibitem [{\citenamefont {Trebst}\ \emph {et~al.}(2007)\citenamefont {Trebst},
  \citenamefont {Werner}, \citenamefont {Troyer}, \citenamefont {Shtengel},\
  and\ \citenamefont {Nayak}}]{Trebst07}%
  \BibitemOpen
  \bibfield  {author} {\bibinfo {author} {\bibfnamefont {S.}~\bibnamefont
  {Trebst}}, \bibinfo {author} {\bibfnamefont {P.}~\bibnamefont {Werner}},
  \bibinfo {author} {\bibfnamefont {M.}~\bibnamefont {Troyer}}, \bibinfo
  {author} {\bibfnamefont {K.}~\bibnamefont {Shtengel}}, \ and\ \bibinfo
  {author} {\bibfnamefont {C.}~\bibnamefont {Nayak}},\ }\href {\doibase
  10.1103/PhysRevLett.98.070602} {\bibfield  {journal} {\bibinfo  {journal}
  {Phys. Rev. Lett.}\ }\textbf {\bibinfo {volume} {98}},\ \bibinfo {pages}
  {070602} (\bibinfo {year} {2007})}\BibitemShut {NoStop}%
\bibitem [{\citenamefont {Vidal}\ \emph {et~al.}(2009)\citenamefont {Vidal},
  \citenamefont {Dusuel},\ and\ \citenamefont {Schmidt}}]{JVidal09a}%
  \BibitemOpen
  \bibfield  {author} {\bibinfo {author} {\bibfnamefont {J.}~\bibnamefont
  {Vidal}}, \bibinfo {author} {\bibfnamefont {S.}~\bibnamefont {Dusuel}}, \
  and\ \bibinfo {author} {\bibfnamefont {K.~P.}\ \bibnamefont {Schmidt}},\
  }\href {\doibase 10.1103/PhysRevB.79.033109} {\bibfield  {journal} {\bibinfo
  {journal} {Phys. Rev. B}\ }\textbf {\bibinfo {volume} {79}},\ \bibinfo
  {pages} {033109} (\bibinfo {year} {2009})}\BibitemShut {NoStop}%
\bibitem [{\citenamefont {Tupitsyn}\ \emph {et~al.}(2010)\citenamefont
  {Tupitsyn}, \citenamefont {Kitaev}, \citenamefont {Prokof'ev},\ and\
  \citenamefont {Stamp}}]{Tupitsyn10}%
  \BibitemOpen
  \bibfield  {author} {\bibinfo {author} {\bibfnamefont {I.~S.}\ \bibnamefont
  {Tupitsyn}}, \bibinfo {author} {\bibfnamefont {A.}~\bibnamefont {Kitaev}},
  \bibinfo {author} {\bibfnamefont {N.~V.}\ \bibnamefont {Prokof'ev}}, \ and\
  \bibinfo {author} {\bibfnamefont {P.~C.~E.}\ \bibnamefont {Stamp}},\ }\href
  {\doibase 10.1103/PhysRevB.82.085114} {\bibfield  {journal} {\bibinfo
  {journal} {Phys. Rev. B}\ }\textbf {\bibinfo {volume} {82}},\ \bibinfo
  {pages} {085114} (\bibinfo {year} {2010})}\BibitemShut {NoStop}%
\bibitem [{\citenamefont {Dusuel}\ \emph {et~al.}(2011)\citenamefont {Dusuel},
  \citenamefont {Kamfor}, \citenamefont {Or\'us}, \citenamefont {Schmidt},\
  and\ \citenamefont {Vidal}}]{Dusuel11}%
  \BibitemOpen
  \bibfield  {author} {\bibinfo {author} {\bibfnamefont {S.}~\bibnamefont
  {Dusuel}}, \bibinfo {author} {\bibfnamefont {M.}~\bibnamefont {Kamfor}},
  \bibinfo {author} {\bibfnamefont {R.}~\bibnamefont {Or\'us}}, \bibinfo
  {author} {\bibfnamefont {K.~P.}\ \bibnamefont {Schmidt}}, \ and\ \bibinfo
  {author} {\bibfnamefont {J.}~\bibnamefont {Vidal}},\ }\href {\doibase
  10.1103/PhysRevLett.106.107203} {\bibfield  {journal} {\bibinfo  {journal}
  {Phys. Rev. Lett.}\ }\textbf {\bibinfo {volume} {106}},\ \bibinfo {pages}
  {107203} (\bibinfo {year} {2011})}\BibitemShut {NoStop}%
\bibitem [{\citenamefont {Wu}\ \emph {et~al.}(2012)\citenamefont {Wu},
  \citenamefont {Deng},\ and\ \citenamefont {Prokof'ev}}]{FWu12}%
  \BibitemOpen
  \bibfield  {author} {\bibinfo {author} {\bibfnamefont {F.}~\bibnamefont
  {Wu}}, \bibinfo {author} {\bibfnamefont {Y.}~\bibnamefont {Deng}}, \ and\
  \bibinfo {author} {\bibfnamefont {N.}~\bibnamefont {Prokof'ev}},\ }\href
  {\doibase 10.1103/PhysRevB.85.195104} {\bibfield  {journal} {\bibinfo
  {journal} {Phys. Rev. B}\ }\textbf {\bibinfo {volume} {85}},\ \bibinfo
  {pages} {195104} (\bibinfo {year} {2012})}\BibitemShut {NoStop}%
\bibitem [{\citenamefont {Schuler}\ \emph {et~al.}(2016)\citenamefont
  {Schuler}, \citenamefont {Whitsitt}, \citenamefont {Henry}, \citenamefont
  {Sachdev},\ and\ \citenamefont {L\"auchli}}]{Schuler16}%
  \BibitemOpen
  \bibfield  {author} {\bibinfo {author} {\bibfnamefont {M.}~\bibnamefont
  {Schuler}}, \bibinfo {author} {\bibfnamefont {S.}~\bibnamefont {Whitsitt}},
  \bibinfo {author} {\bibfnamefont {L.-P.}\ \bibnamefont {Henry}}, \bibinfo
  {author} {\bibfnamefont {S.}~\bibnamefont {Sachdev}}, \ and\ \bibinfo
  {author} {\bibfnamefont {A.~M.}\ \bibnamefont {L\"auchli}},\ }\href {\doibase
  10.1103/PhysRevLett.117.210401} {\bibfield  {journal} {\bibinfo  {journal}
  {Phys. Rev. Lett.}\ }\textbf {\bibinfo {volume} {117}},\ \bibinfo {pages}
  {210401} (\bibinfo {year} {2016})}\BibitemShut {NoStop}%
\bibitem [{\citenamefont {Kitaev}(2006)}]{Kitaev06}%
  \BibitemOpen
  \bibfield  {author} {\bibinfo {author} {\bibfnamefont {A.}~\bibnamefont
  {Kitaev}},\ }\href {\doibase 10.1016/j.aop.2005.10.005} {\bibfield  {journal}
  {\bibinfo  {journal} {Ann. Phys.}\ }\textbf {\bibinfo {volume} {321}},\
  \bibinfo {pages} {2 } (\bibinfo {year} {2006})}\BibitemShut {NoStop}%
\bibitem [{\citenamefont {Levin}\ and\ \citenamefont {Wen}(2005)}]{Levin05}%
  \BibitemOpen
  \bibfield  {author} {\bibinfo {author} {\bibfnamefont {M.~A.}\ \bibnamefont
  {Levin}}\ and\ \bibinfo {author} {\bibfnamefont {X.-G.}\ \bibnamefont
  {Wen}},\ }\href {\doibase 10.1103/PhysRevB.71.045110} {\bibfield  {journal}
  {\bibinfo  {journal} {Phys. Rev. B}\ }\textbf {\bibinfo {volume} {71}},\
  \bibinfo {pages} {045110} (\bibinfo {year} {2005})}\BibitemShut {NoStop}%
\bibitem [{\citenamefont {Jackeli}\ and\ \citenamefont
  {Khaliullin}(2009)}]{Jackeli09}%
  \BibitemOpen
  \bibfield  {author} {\bibinfo {author} {\bibfnamefont {G.}~\bibnamefont
  {Jackeli}}\ and\ \bibinfo {author} {\bibfnamefont {G.}~\bibnamefont
  {Khaliullin}},\ }\href {\doibase 10.1103/PhysRevLett.102.017205} {\bibfield
  {journal} {\bibinfo  {journal} {Phys. Rev. Lett.}\ }\textbf {\bibinfo
  {volume} {102}},\ \bibinfo {pages} {017205} (\bibinfo {year}
  {2009})}\BibitemShut {NoStop}%
\bibitem [{\citenamefont {Chaloupka}\ \emph {et~al.}(2010)\citenamefont
  {Chaloupka}, \citenamefont {Jackeli},\ and\ \citenamefont
  {Khaliullin}}]{Chaloupka10}%
  \BibitemOpen
  \bibfield  {author} {\bibinfo {author} {\bibfnamefont {J.}~\bibnamefont
  {Chaloupka}}, \bibinfo {author} {\bibfnamefont {G.}~\bibnamefont {Jackeli}},
  \ and\ \bibinfo {author} {\bibfnamefont {G.}~\bibnamefont {Khaliullin}},\
  }\href {\doibase 10.1103/PhysRevLett.105.027204} {\bibfield  {journal}
  {\bibinfo  {journal} {Phys. Rev. Lett.}\ }\textbf {\bibinfo {volume} {105}},\
  \bibinfo {pages} {027204} (\bibinfo {year} {2010})}\BibitemShut {NoStop}%
\bibitem [{\citenamefont {Jian}\ and\ \citenamefont {Qi}(2014)}]{CMJian14}%
  \BibitemOpen
  \bibfield  {author} {\bibinfo {author} {\bibfnamefont {C.-M.}\ \bibnamefont
  {Jian}}\ and\ \bibinfo {author} {\bibfnamefont {X.-L.}\ \bibnamefont {Qi}},\
  }\href {\doibase 10.1103/PhysRevX.4.041043} {\bibfield  {journal} {\bibinfo
  {journal} {Phys. Rev. X}\ }\textbf {\bibinfo {volume} {4}},\ \bibinfo {pages}
  {041043} (\bibinfo {year} {2014})}\BibitemShut {NoStop}%
\bibitem [{\citenamefont {Fuji}\ and\ \citenamefont {Furusaki}(2019)}]{Fuji19}%
  \BibitemOpen
  \bibfield  {author} {\bibinfo {author} {\bibfnamefont {Y.}~\bibnamefont
  {Fuji}}\ and\ \bibinfo {author} {\bibfnamefont {A.}~\bibnamefont
  {Furusaki}},\ }\href {\doibase 10.1103/PhysRevB.99.241107} {\bibfield
  {journal} {\bibinfo  {journal} {Phys. Rev. B}\ }\textbf {\bibinfo {volume}
  {99}},\ \bibinfo {pages} {241107} (\bibinfo {year} {2019})}\BibitemShut
  {NoStop}%
\bibitem [{\citenamefont {Iadecola}\ \emph {et~al.}(2019)\citenamefont
  {Iadecola}, \citenamefont {Neupert}, \citenamefont {Chamon},\ and\
  \citenamefont {Mudry}}]{Iadecola19}%
  \BibitemOpen
  \bibfield  {author} {\bibinfo {author} {\bibfnamefont {T.}~\bibnamefont
  {Iadecola}}, \bibinfo {author} {\bibfnamefont {T.}~\bibnamefont {Neupert}},
  \bibinfo {author} {\bibfnamefont {C.}~\bibnamefont {Chamon}}, \ and\ \bibinfo
  {author} {\bibfnamefont {C.}~\bibnamefont {Mudry}},\ }\href {\doibase
  10.1103/PhysRevB.99.245138} {\bibfield  {journal} {\bibinfo  {journal} {Phys.
  Rev. B}\ }\textbf {\bibinfo {volume} {99}},\ \bibinfo {pages} {245138}
  (\bibinfo {year} {2019})}\BibitemShut {NoStop}%
\end{thebibliography}%

\end{document}